\documentclass{aastex631}
\usepackage{savesym}
\savesymbol{tablenum}
\usepackage{siunitx}
\restoresymbol{SIX}{tablenum}

\graphicspath{{./}{figures/}}

\begin{document}

\title{IceCat-1: the IceCube Event Catalog of Alert Tracks}

\affiliation{III. Physikalisches Institut, RWTH Aachen University, D-52056 Aachen, Germany}
\affiliation{Department of Physics, University of Adelaide, Adelaide, 5005, Australia}
\affiliation{Dept. of Physics and Astronomy, University of Alaska Anchorage, 3211 Providence Dr., Anchorage, AK 99508, USA}
\affiliation{Dept. of Physics, University of Texas at Arlington, 502 Yates St., Science Hall Rm 108, Box 19059, Arlington, TX 76019, USA}
\affiliation{CTSPS, Clark-Atlanta University, Atlanta, GA 30314, USA}
\affiliation{School of Physics and Center for Relativistic Astrophysics, Georgia Institute of Technology, Atlanta, GA 30332, USA}
\affiliation{Dept. of Physics, Southern University, Baton Rouge, LA 70813, USA}
\affiliation{Dept. of Physics, University of California, Berkeley, CA 94720, USA}
\affiliation{Lawrence Berkeley National Laboratory, Berkeley, CA 94720, USA}
\affiliation{Institut f{\"u}r Physik, Humboldt-Universit{\"a}t zu Berlin, D-12489 Berlin, Germany}
\affiliation{Fakult{\"a}t f{\"u}r Physik {\&} Astronomie, Ruhr-Universit{\"a}t Bochum, D-44780 Bochum, Germany}
\affiliation{Universit{\'e} Libre de Bruxelles, Science Faculty CP230, B-1050 Brussels, Belgium}
\affiliation{Vrije Universiteit Brussel (VUB), Dienst ELEM, B-1050 Brussels, Belgium}
\affiliation{Department of Physics and Laboratory for Particle Physics and Cosmology, Harvard University, Cambridge, MA 02138, USA}
\affiliation{Dept. of Physics, Massachusetts Institute of Technology, Cambridge, MA 02139, USA}
\affiliation{Dept. of Physics and The International Center for Hadron Astrophysics, Chiba University, Chiba 263-8522, Japan}
\affiliation{Department of Physics, Loyola University Chicago, Chicago, IL 60660, USA}
\affiliation{Dept. of Physics and Astronomy, University of Canterbury, Private Bag 4800, Christchurch, New Zealand}
\affiliation{Dept. of Physics, University of Maryland, College Park, MD 20742, USA}
\affiliation{Dept. of Astronomy, Ohio State University, Columbus, OH 43210, USA}
\affiliation{Dept. of Physics and Center for Cosmology and Astro-Particle Physics, Ohio State University, Columbus, OH 43210, USA}
\affiliation{Niels Bohr Institute, University of Copenhagen, DK-2100 Copenhagen, Denmark}
\affiliation{Dept. of Physics, TU Dortmund University, D-44221 Dortmund, Germany}
\affiliation{Dept. of Physics and Astronomy, Michigan State University, East Lansing, MI 48824, USA}
\affiliation{Dept. of Physics, University of Alberta, Edmonton, Alberta, Canada T6G 2E1}
\affiliation{Erlangen Centre for Astroparticle Physics, Friedrich-Alexander-Universit{\"a}t Erlangen-N{\"u}rnberg, D-91058 Erlangen, Germany}
\affiliation{Physik-Department, Technische Universit{\"a}t M{\"u}nchen, D-85748 Garching, Germany}
\affiliation{D{\'e}partement de physique nucl{\'e}aire et corpusculaire, Universit{\'e} de Gen{\`e}ve, CH-1211 Gen{\`e}ve, Switzerland}
\affiliation{Dept. of Physics and Astronomy, University of Gent, B-9000 Gent, Belgium}
\affiliation{Dept. of Physics and Astronomy, University of California, Irvine, CA 92697, USA}
\affiliation{Karlsruhe Institute of Technology, Institute for Astroparticle Physics, D-76021 Karlsruhe, Germany }
\affiliation{Karlsruhe Institute of Technology, Institute of Experimental Particle Physics, D-76021 Karlsruhe, Germany }
\affiliation{Dept. of Physics, Engineering Physics, and Astronomy, Queen's University, Kingston, ON K7L 3N6, Canada}
\affiliation{Department of Physics {\&} Astronomy, University of Nevada, Las Vegas, NV, 89154, USA}
\affiliation{Nevada Center for Astrophysics, University of Nevada, Las Vegas, NV 89154, USA}
\affiliation{Dept. of Physics and Astronomy, University of Kansas, Lawrence, KS 66045, USA}
\affiliation{Department of Physics and Astronomy, UCLA, Los Angeles, CA 90095, USA}
\affiliation{Centre for Cosmology, Particle Physics and Phenomenology - CP3, Universit{\'e} catholique de Louvain, Louvain-la-Neuve, Belgium}
\affiliation{Department of Physics, Mercer University, Macon, GA 31207-0001, USA}
\affiliation{Dept. of Astronomy, University of Wisconsin{\textendash}Madison, Madison, WI 53706, USA}
\affiliation{Dept. of Physics and Wisconsin IceCube Particle Astrophysics Center, University of Wisconsin{\textendash}Madison, Madison, WI 53706, USA}
\affiliation{Institute of Physics, University of Mainz, Staudinger Weg 7, D-55099 Mainz, Germany}
\affiliation{Department of Physics, Marquette University, Milwaukee, WI, 53201, USA}
\affiliation{Institut f{\"u}r Kernphysik, Westf{\"a}lische Wilhelms-Universit{\"a}t M{\"u}nster, D-48149 M{\"u}nster, Germany}
\affiliation{Bartol Research Institute and Dept. of Physics and Astronomy, University of Delaware, Newark, DE 19716, USA}
\affiliation{Dept. of Physics, Yale University, New Haven, CT 06520, USA}
\affiliation{Columbia Astrophysics and Nevis Laboratories, Columbia University, New York, NY 10027, USA}
\affiliation{Dept. of Physics, University of Oxford, Parks Road, Oxford OX1 3PU, UK}
\affiliation{Dipartimento di Fisica e Astronomia Galileo Galilei, Universit{\`a} Degli Studi di Padova, 35122 Padova PD, Italy}
\affiliation{Dept. of Physics, Drexel University, 3141 Chestnut Street, Philadelphia, PA 19104, USA}
\affiliation{Physics Department, South Dakota School of Mines and Technology, Rapid City, SD 57701, USA}
\affiliation{Dept. of Physics, University of Wisconsin, River Falls, WI 54022, USA}
\affiliation{Dept. of Physics and Astronomy, University of Rochester, Rochester, NY 14627, USA}
\affiliation{Department of Physics and Astronomy, University of Utah, Salt Lake City, UT 84112, USA}
\affiliation{Oskar Klein Centre and Dept. of Physics, Stockholm University, SE-10691 Stockholm, Sweden}
\affiliation{Dept. of Physics and Astronomy, Stony Brook University, Stony Brook, NY 11794-3800, USA}
\affiliation{Dept. of Physics, Sungkyunkwan University, Suwon 16419, Korea}
\affiliation{Institute of Physics, Academia Sinica, Taipei, 11529, Taiwan}
\affiliation{Dept. of Physics and Astronomy, University of Alabama, Tuscaloosa, AL 35487, USA}
\affiliation{Dept. of Astronomy and Astrophysics, Pennsylvania State University, University Park, PA 16802, USA}
\affiliation{Dept. of Physics, Pennsylvania State University, University Park, PA 16802, USA}
\affiliation{Dept. of Physics and Astronomy, Uppsala University, Box 516, S-75120 Uppsala, Sweden}
\affiliation{Dept. of Physics, University of Wuppertal, D-42119 Wuppertal, Germany}
\affiliation{Deutsches Elektronen-Synchrotron DESY, Platanenallee 6, 15738 Zeuthen, Germany }

\author[0000-0001-6141-4205]{R. Abbasi}
\affiliation{Department of Physics, Loyola University Chicago, Chicago, IL 60660, USA}

\author[0000-0001-8952-588X]{M. Ackermann}
\affiliation{Deutsches Elektronen-Synchrotron DESY, Platanenallee 6, 15738 Zeuthen, Germany }

\author{J. Adams}
\affiliation{Dept. of Physics and Astronomy, University of Canterbury, Private Bag 4800, Christchurch, New Zealand}

\author[0000-0002-9714-8866]{S. K. Agarwalla}
\altaffiliation{also at Institute of Physics, Sachivalaya Marg, Sainik School Post, Bhubaneswar 751005, India}
\affiliation{Dept. of Physics and Wisconsin IceCube Particle Astrophysics Center, University of Wisconsin{\textendash}Madison, Madison, WI 53706, USA}

\author[0000-0003-2252-9514]{J. A. Aguilar}
\affiliation{Universit{\'e} Libre de Bruxelles, Science Faculty CP230, B-1050 Brussels, Belgium}

\author[0000-0003-0709-5631]{M. Ahlers}
\affiliation{Niels Bohr Institute, University of Copenhagen, DK-2100 Copenhagen, Denmark}

\author[0000-0002-9534-9189]{J.M. Alameddine}
\affiliation{Dept. of Physics, TU Dortmund University, D-44221 Dortmund, Germany}

\author{N. M. Amin}
\affiliation{Bartol Research Institute and Dept. of Physics and Astronomy, University of Delaware, Newark, DE 19716, USA}

\author{K. Andeen}
\affiliation{Department of Physics, Marquette University, Milwaukee, WI, 53201, USA}

\author[0000-0003-2039-4724]{G. Anton}
\affiliation{Erlangen Centre for Astroparticle Physics, Friedrich-Alexander-Universit{\"a}t Erlangen-N{\"u}rnberg, D-91058 Erlangen, Germany}

% ...

\begin{abstract}
We present a catalog of likely astrophysical neutrino track-like events from the IceCube Neutrino Observatory. IceCube began reporting likely astrophysical neutrinos in 2016 and this system was updated in 2019. The catalog presented here includes events that were reported in real-time since 2019, as well as events identified in archival data samples starting from 2011. We report 275 neutrino events from two selection channels as the first entries in the catalog, the IceCube Event Catalog of Alert Tracks,  which will see ongoing extensions with additional alerts.  The Gold and Bronze alert channels respectively provide neutrino candidates with 50\% and 30\% probability of being astrophysical, on average assuming an astrophysical neutrino power law energy spectral index of 2.19. For each neutrino alert, we provide the reconstructed energy, direction, false alarm rate, probability of being astrophysical in origin, and likelihood contours describing the spatial uncertainty in the alert’s reconstructed location. We also investigate a directional correlation of these neutrino events with gamma-ray and X-ray catalogs including 4FGL, 3HWC, TeVCat and Swift-BAT. 
    
\end{abstract}


\section{Introduction}\label{sec:intro}

The emerging field of multimessenger astronomy combines measurements taken across the electromagnetic spectrum with neutrinos and gravitational waves to elucidate the nature of astrophysical objects. Notable examples of discoveries in recent years include the joint gravitational wave and electromagnetic observation of a binary neutron star merger \citep{LIGOScientific:2017ync}, and the coincident detection of neutrinos and gamma rays from the blazar TXS 0506+056 \citep{txs}. Breakthroughs like the latter hold the key to identifying the sites of hadronic acceleration and solving a major open puzzle in modern astrophysics, the origin of cosmic rays. Astrophysical neutrinos are produced in either $pp$ collisions or $p\gamma$ interactions following cosmic-ray acceleration, and neutrino observatories like IceCube have a unique role in probing the distant universe in the TeV--PeV energy regime. The prompt observation of transient phenomena, such as gamma-ray bursts, tidal disruption events, and supernovae, in different wavelengths and messengers requires the rapid sharing of information between different observational facilities. Since 2016, IceCube has been issuing real-time alerts within minutes of the detection of astrophysical neutrino candidates \citep{IceCube:2016zyt}. Several improvements were introduced in the real-time stream in 2019 \citep{Blaufuss:2019fgv}. The updated program includes increased signal purity, better rejection of backgrounds, and an expanded alert selection resulting in more frequent alerts from IceCube than the previous alert program. These improvements also introduced a two-level classification of signal purity in the form of ``Gold'' and ``Bronze'' alerts. 
 In this work, we describe the improvements made to the real-time alert selection, apply the updated selection to archival IceCube data going back to 2011 when IceCube first began full operations with 86 strings, and present the first catalog of neutrino events of likely astrophysical origin. This catalog, the IceCube Event Catalog of Alert Tracks (ICECAT-1), contains detailed information on key parameters of 275 neutrino events detected between 13 May 2011 and 31 December 2020, providing a unique sample for  multimessenger studies. The accompanying data release provides the log-likelihood sky-maps and spatial uncertainties for all events. This will also establish a framework for continued data releases from future alerts, including additions from the most recent IceCube alerts.

This paper is structured as follows. We introduce the IceCube Neutrino Observatory and real-time data selection for this catalog in Section \ref{sec:detector}. In Section \ref{sec:alerts}, we describe how the alert events are further processed and prepared for follow-ups. We discuss the overall properties of the catalog in Section \ref{sec:cat}. We describe a potential search for correlations with a few multi-wavelength catalogs in Section \ref{sec:assc}, and conclude in section \ref{sec:conclusion}.


\section{Detector and Event Selection}\label{sec:detector}
The IceCube Neutrino Observatory consists of 86 strings of photo-detectors embedded in a cubic kilometer of ice beneath the South Pole. The photo-detectors, known as digital optical modules (DOMs), are spaced along the vertical length of each string \citep{icecube_daq}. The strings are arranged on average 125 m apart in a hexagonal grid with a more densely instrumented set of strings located in the center of the array is known as  DeepCore \citep{IceCube:2011ucd}. In addition to the in-ice detectors, there is also a surface array of 162 ice-filled tanks instrumented with two DOMs each, known as IceTop \citep{Abbasi_2013}. The surface array functions as a detector for air showers induced by cosmic rays and gamma rays.   

IceCube detects the Cherenkov light produced by the secondary charged particles from neutrino interactions propagating through the ice. The total number of photo-electrons (PE) detected (deposited charge) and their arrival times are used to reconstruct the deposited energy and the incoming direction of these charged particles. The optical emission signatures can be classified into two distinct types of event morphologies: tracks and  cascades. Track-like events are predominantly produced by muons, which originate in charge-current (CC) interactions of muon neutrinos and from cosmic-ray-induced showers. At final selection level, the majority of muon track-like events detected pass fully through the instrumented volume; however, tracks starting or stopping within the instrumented volume are observed.  Starting tracks in particular, generated by a muon neutrino CC interaction within the IceCube instrumented volume can be a strong indication of astrophysical origin \cite{IceCube:2020wum}.  The directions of such events can be reconstructed with an uncertainty of less than 1$^\circ$ \cite{IceCube:2013dkx}. The location of the neutrino interaction, which can be $\mathcal{O}$(km) outside the instrumented volume, and the length of the track captured within the detector  can lead to  large uncertainties in the  measured neutrino energy. Cascades are produced by all-flavor neutral-current (NC) neutrino interactions as well as electron-neutrino CC interactions. These events deposit all their energy within spherical showers of $\mathcal{O}(10)$ m and can only be resolved with angular uncertainties of $\sim 10^\circ$ ~\citep{IceCube:2013dkx,Abbasi:2021ryj}. Additionally, cascade-like signatures can arise from charge-current interactions of tau neutrinos \cite{IceCube:2020acn} or from neutrinos at the Glashow resonance \cite{IceCube:2021rpz}.  Tracks -- due to their superior angular resolution --  are best suited for use in multimessenger searches for astrophysical sources, and are the primary constituents of IceCube real-time alerts. In July 2020, IceCube also began issuing alerts for cascade-like events\footnote{\url{https://gcn.gsfc.nasa.gov/amon_icecube_cascade_events.html}}. However, the cascade sample is not part of the catalog in this paper.

\subsection{Real-time Reconstruction and Communication}
IceCube detects neutrinos at a rate of a few mHz, the vast majority of which are atmospheric neutrinos produced in cosmic-ray interactions in the Earth's atmosphere \citep{IceCube:2010whx,IceCube:2014slq}. A real-time infrastructure identifies events with a significant probability of being of astrophysical origin (neutrino energy $>\sim$100~TeV) and  promptly alerts the astronomy community to such a detection. The system is described in detail in \cite{realtime}. We discuss it briefly here with an emphasis on updates and improvements relevant for this catalog. 

An online filtering system at the South Pole identifies candidate neutrino events. The candidate events are reconstructed on several hundred parallel filter clients to determine their observed energies, directions, and morphology. Additional selection criteria (\cite{realtime}) are applied to determine whether an event passes the preliminary online alert criteria. A single online alert writer process collates these events for transmission to I3Live -- the IceCube experiment control system \citep{IceCube:2016zyt}. The key event summary data for candidate events passing the quality cuts are relayed to the IceCube data center in the Northern hemisphere over satellite. The full event information including the signals registered by the DOMs follows in a second message that is used for a detailed follow-up reconstructions as explained in section \ref{sec:alerts}.

A dedicated computing system, located at the IceCube data center in the Northern hemisphere,  further evaluates the alert candidates arriving from the South Pole to check if they pass the online alert criteria. If selected, an alert message is generated and distributed to the public through the Astrophysical
Multimessenger Observatory Network (AMON) system \citep{AyalaSolares:2019iiy} which utilizes the General Coordinates Network\footnote{\url{https://gcn.nasa.gov/}}  (GCN) for communication. The whole chain of events, from the neutrino detection to the issuance of an alert is fully automated and takes between 30 s to 40 s on average.

\subsection{Updated Event Selections}
The updated event selection introduced in 2019 includes two key improvements that aim to convey the detection of potential astrophysical tracks to the community as frequently as possible. One is the introduction of ``Gold'' and ``Bronze'' streams that classify alerts based on their likelihood of being astrophysical in nature. The classification is based on a quantity called ``signalness,'' which is defined as,

\begin{equation}
   \text{Signalness} (E,\delta) = \frac{N_{\text{signal}}(E,\delta)}{N_{\rm{signal}}(E,\delta) + N_{\rm{background}}(E,\delta)}. 
\end{equation}
Here $E$ is the reconstructed event energy, $\delta$ is the event declination, $N_{\rm{signal}} (E,\delta)$ and $N_{\rm{background}} (E,\delta)$ are the expected number of signal and background events at declination $\delta$ and above energy $E$ determined from simulations. $N_{\rm{signal}} (E,\delta)$ and $N_{\rm{background}} (E,\delta)$, and therefore signalness are functions of the assumed astrophysical neutrino spectrum. The streams were optimized on simulations using an astrophysical neutrino spectrum of $E^{-2.19}$ \citep{Haack:2017E1}. The signalness quantity assigns a probability of being an astrophysical neutrino to each alert event, assuming the same $E^{-2.19}$ astrophysical neutrino spectrum.  The alert generation criteria are optimized such that ``Bronze'' alerts have have an average signalness value between 30 \% and 50 \%, whereas ``Gold'' candidates have an average signalness above 50\%. Thus, the Gold stream has a higher signal purity. We note that the signalness is calculated after event selection to grade alerts, and is not explicitly used in alert selection. Certain tracks with high signalness values may not pass the real-time criteria and would end up in other selections. The signalness of each alert is sent out as part of the GCN notice. The ``notice type'' field indicates the Bronze (Gold) alerts as ICECUBE Astrotrack Bronze (Gold). An example of a GCN notice for IC190730A can be seen at the following URL \url{https://gcn.gsfc.nasa.gov/notices_amon_g_b/132910_57145925.amon}.   

The second improvement to the updated event selection is the introduction of a new track selection known as Gamma-ray Follow-Up (GFU) selection. This complements the previously existing selections in the real-time scheme. The event selections are summarized below.

\begin{figure}[tbp!]
\includegraphics[width=0.44\textwidth, trim=0cm 0cm 0cm 0cm, clip=true]{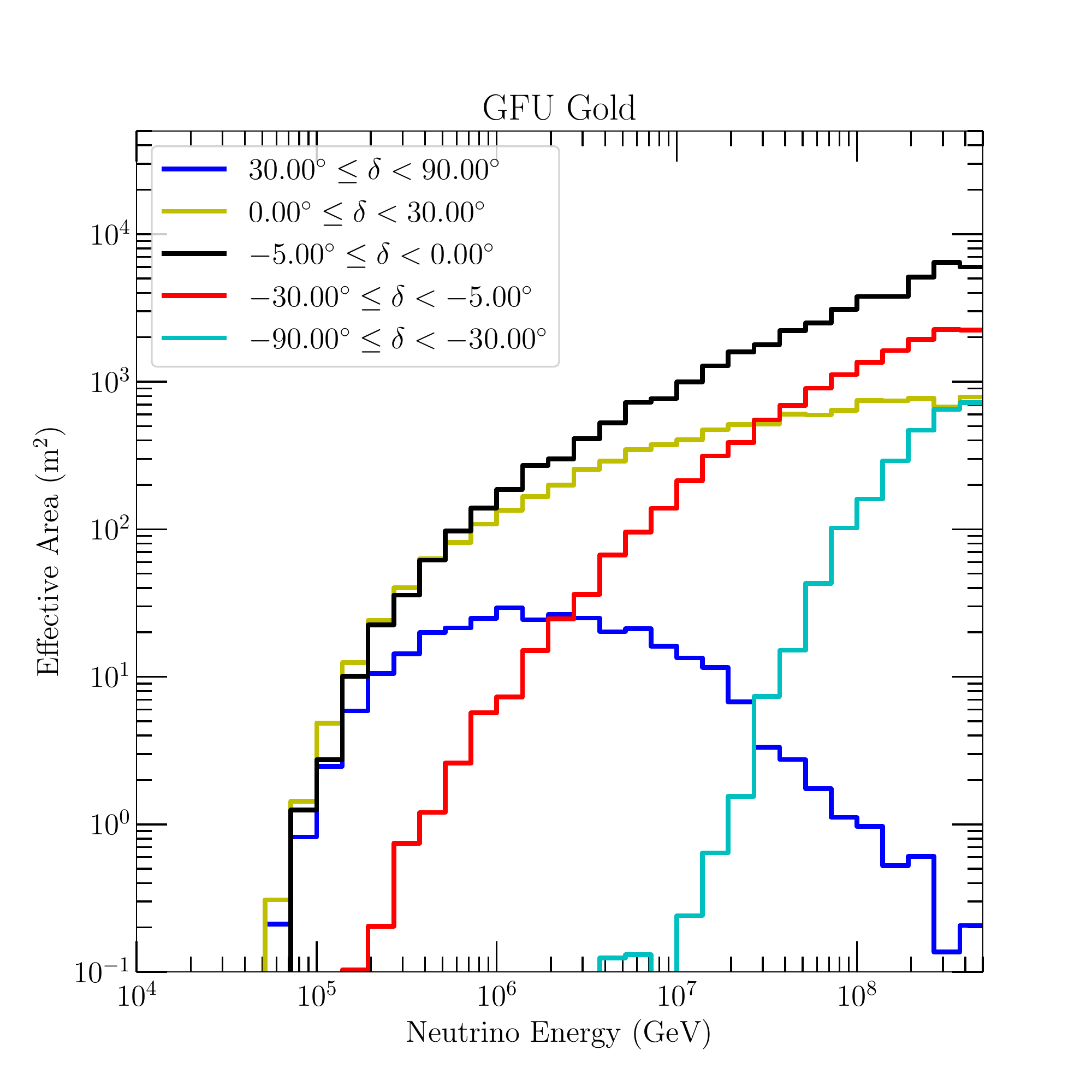}\includegraphics[width=0.44\textwidth, trim=0cm 0cm 0cm 0cm, clip=true]{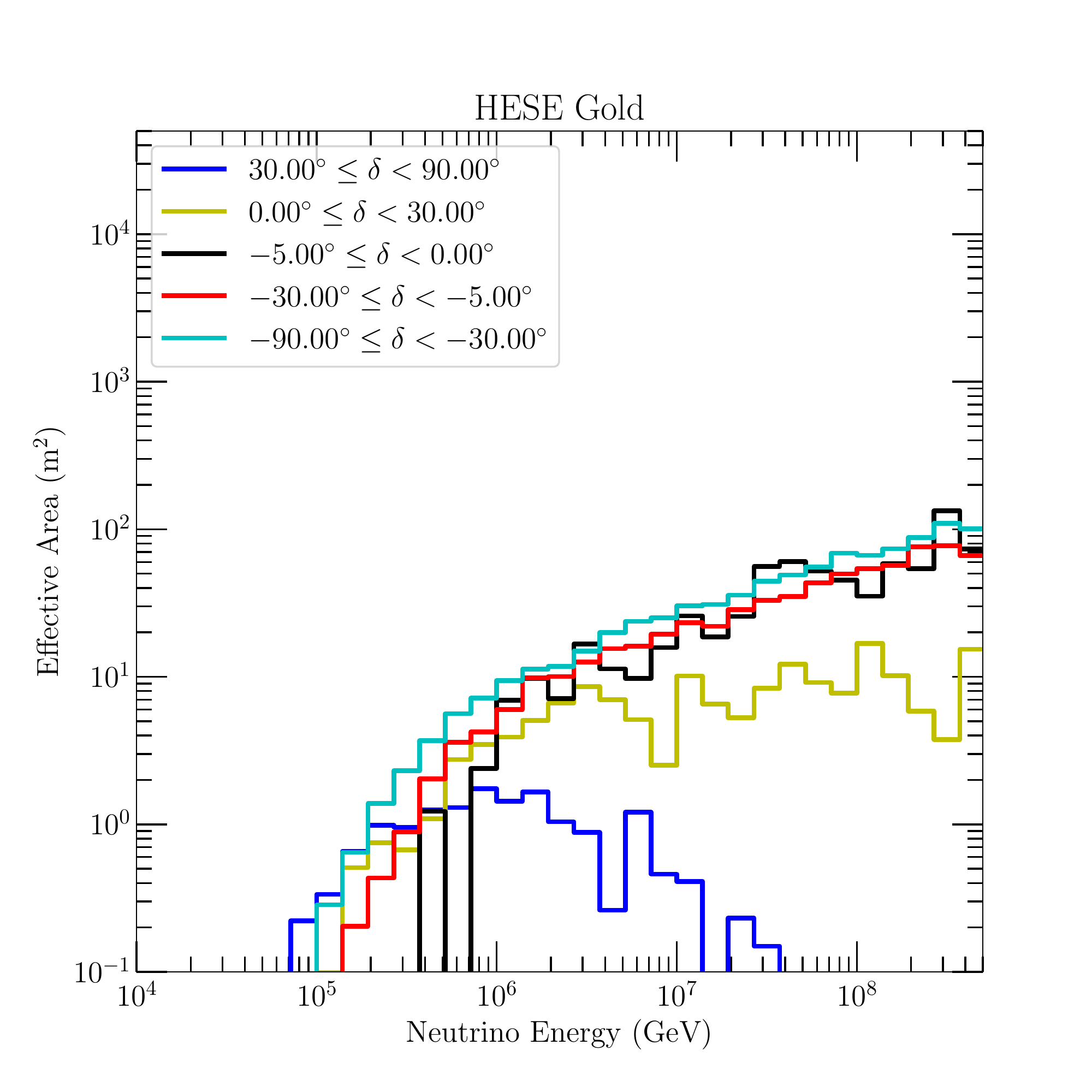} \\ \includegraphics[width=0.44\textwidth, trim=0cm 0cm 0cm 0cm, clip=true]{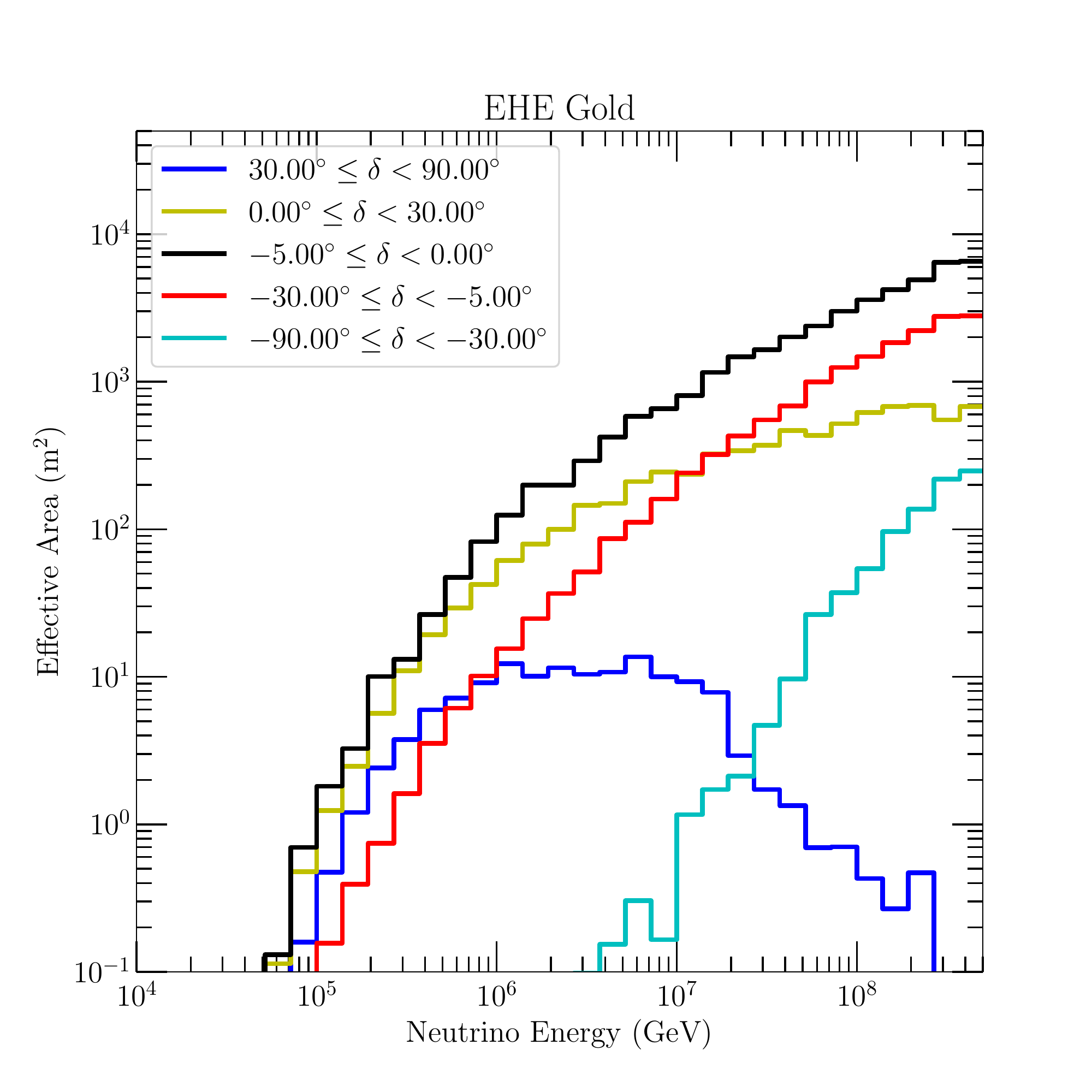}
\includegraphics[width=0.44\textwidth, trim=0cm 0cm 0cm 0cm, clip=true]{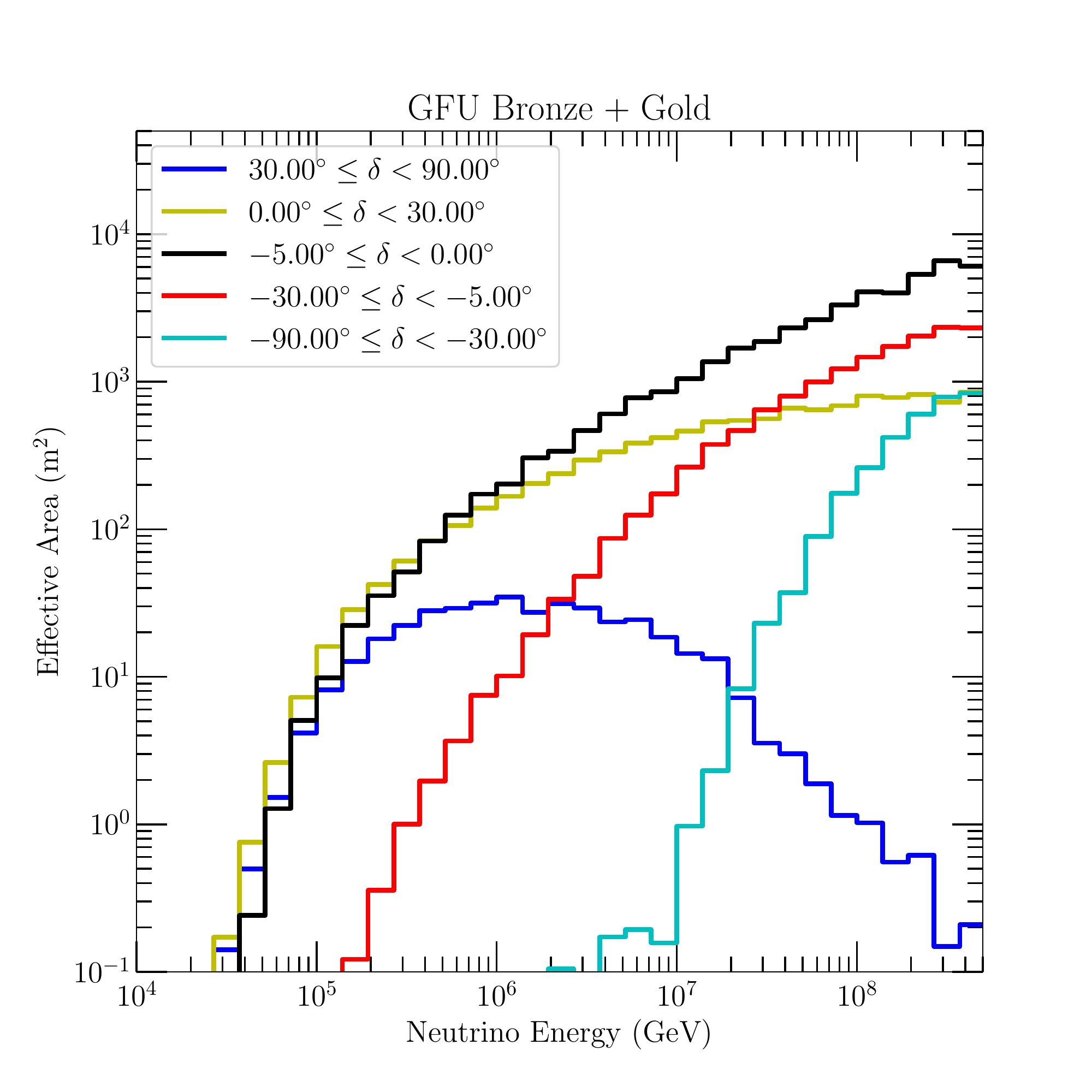}\\ \includegraphics[width=0.44\textwidth, trim=0cm 0cm 0cm 0cm, clip=true]{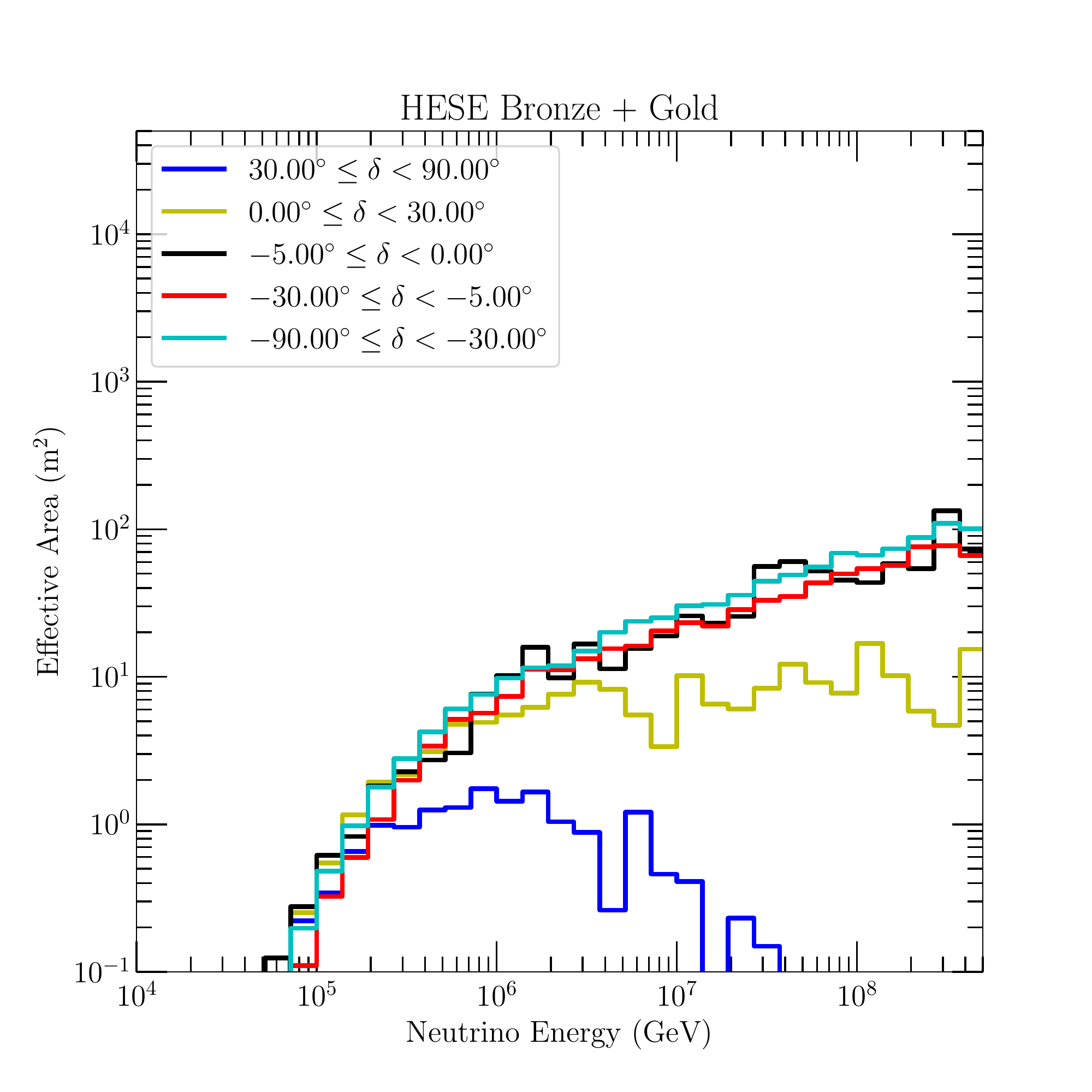}
\caption{The $\nu_{\mu}+\bar{\nu_{\mu}}$ effective areas for the event selections used in this catalog as a function of neutrino energy. The solid lines show the
effective area in different declination bands: [$30^\circ$,$90^\circ$] (blue), [$0^\circ$,$30^\circ$] (yellow), [$-5^\circ$,$0^\circ$] (black), [$-30^\circ$,$-5^\circ$] (red) and [$-90^\circ$,$-30^\circ$] (cyan).}
\label{fig:eA}

\end{figure}


\subsubsection{Gamma-Ray Followup Event Selection}
This event selection is based on an existing IceCube data neutrino candidate selection that was originally designed to provide triggers for follow-up by Imaging Air Cherenkov Telescopes in gamma rays -- hence the name Gamma-ray Followup or GFU \citep{IceCube:2016xci}. This reconstruction targets through-going tracks and employs separate boosted decision tree-based selections for events from the Northern and Southern hemisphere (up-going and down-going events in the IceCube reference frame, respectively) to suppress atmospheric backgrounds. A threshold is applied to the reconstructed event energy to achieve the 30\% and 50\% signalness criteria for alerts. This results in only the highest energy events (hundreds of TeV) being selected. Figure \ref{fig:eA} shows the effective area for the GFU Gold and Bronze alert selection as a function of neutrino energy. The majority (86\%) of the alerts issued by IceCube fall under the GFU selection. The ten-year catalog includes 72 GFU Gold and 164 GFU Bronze events. 

\clearpage

\subsubsection{High-Energy Starting Event Selection}
The High-Energy Starting Event (HESE) selection for alerts includes only starting tracks,  track-like events that have the neutrino interaction vertex inside the fiducial volume of the detector \citep{IceCube:2020wum}. This technique efficiently rejects the atmospheric muon background \citep{IceCube:2013low}. Since, the highest energy events are more likely to be of astrophysical origin, only the events that have a total deposited charge in the detector of at least 6000 photoelectrons are considered. As an improvement to previous HESE alerts~\cite{realtime}, additional cuts are introduced to further reduce poorly reconstructed event. We only use events that have a minimum measured track length of 200 m. Due to the effective veto requirement for HESE alerts,  down-going events from the Southern hemisphere can be observed at lower neutrino energies, as illustrated by the HESE effective area for different declination bins in Fig. \ref{fig:eA}. The ten-year catalog includes 9 HESE Gold and 8 HESE Bronze events.

\subsubsection{Extremely High-Energy Event Selection}
The Extremely High-Energy Event (EHE) selection is optimized for detecting track-like neutrino events with energies between 500 TeV and 10 PeV. Atmospheric backgrounds are minimized by employing a two dimensional cut in the plane of the reconstructed zenith angle and the logarithm of the deposited charge detected. The selection is unchanged from the one described in \cite{realtime}, and the cuts are set to achieve an average signalness of 50\%, and therefore EHE events are only sent as part of the Gold stream.  Figure \ref{fig:eA} shows the effective area for the EHE selection as a function of neutrino energy.  This catalog contains 22 events that passed the EHE selection during the 9.6-year period.

\subsection{Expected and Observed Rates}
\begin{figure}[hb!]
\includegraphics[width=1\textwidth]{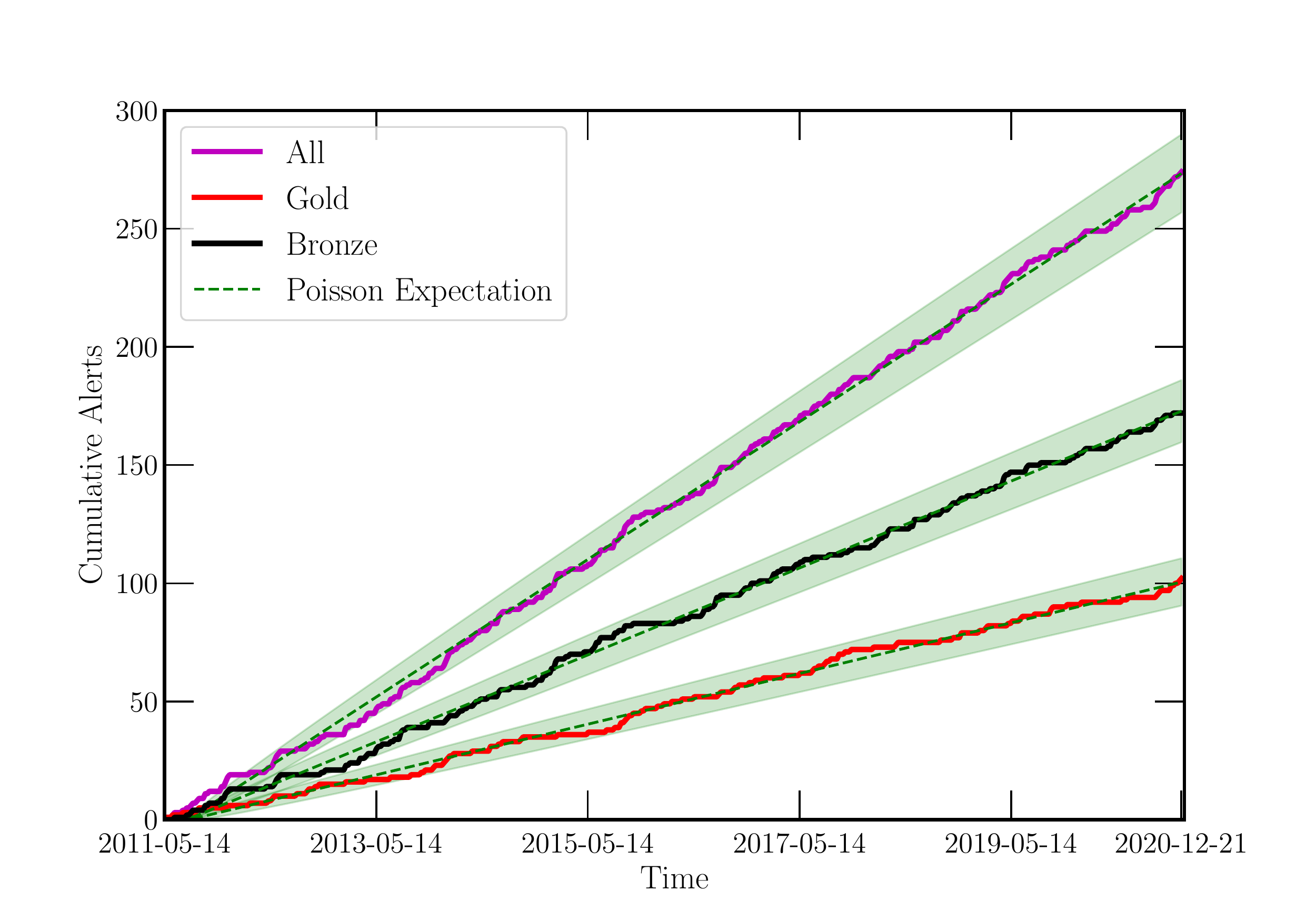}
\caption{The cumulative number of alerts as a function of time. The solid
black and red lines show the number of Gold and Bronze alerts respectively, while
the magenta line shows the combined number for all alerts. The dashed green
line and the shaded green band show the median and standard deviation of the
best-fit Poisson distribution to the number of alerts in each category.}
\label{fig:cumulative}
\end{figure}

Figure \ref{fig:cumulative} shows the time-evolution of the number of observed alerts over the years. The cumulative number of total alerts,  as a function of year are best described by a straight line with slope $28.6$ alerts per year.  Table \ref{tab:rates} shows the number of expected and observed number of events in the 3,514 days of the IceCube data used in this work.  We calculate the expected number of alerts arising from astrophysical sources for each selection by multiplying its respective effective area with IceCube's latest measured diffuse astrophysical muon neutrino spectrum \citep{IceCube:2021uhz} that reports a spectral power-law index of 2.37. Since, the event selection was originally optimized assuming a spectral index of 2.19 based on a previous IceCube measurement \citep{Haack:2017E1}, the numbers reported here slightly differ from the ones in \cite{Blaufuss:2019fgv}. IceCube continues to take data and refine its reconstruction methods resulting in a more precise measurement of the astrophysical neutrino spectrum over the years. This evolution is reflected in the use of an updated spectral index in this work. The expected number of signal events can change up to $\sim 15\%$ when using a spectral index of 2.19 instead of 2.37. The expected number of background events is calculated using a simulation of atmospheric muons and neutrinos \citep{PhysRevD.79.043009,Heck:1998vt}. We also note that the event selections are not mutually exclusive -- a single event may pass multiple selections. In particular, GFU and EHE selections have significant overlap as they both focus on through-going tracks. The expected numbers in Table \ref{tab:rates}, account for the overlap and only report the unique events from each stream. In real-time, if an event passes multiple selections only one alert is issued based on a hierarchical rule for labeling. The hierarchical scheme in order of preference for the Gold stream is GFU Gold, EHE Gold and HESE Gold. Similarly, for the Bronze stream, a GFU Bronze alert is sent preferentially over a HESE Bronze alert. An event passing both the Gold and Bronze selections is only sent in the Gold stream. The preference order is decided based on the angular resolutions and relative signal purity of the different streams. 

The observed rates of alerts from the Gold and Bronze channels shown in Table \ref{tab:rates} are compatible with expectations when considering Poisson fluctuations as well as the uncertainties in the astrophysical neutrino spectral parameters and the background modeling. For instance, considering the errors on the measured diffuse muon neutrino spectral parameters \citep{IceCube:2021uhz} , the expected number of GFU Gold signal events can be as low as 39.1, bringing the overall expected rate from 101.3 to 86.1, which is within $\sim 1.5 \sigma$ of the observed number of 72 events.  It should be noted that while the average signalness of the Gold and Bronze selection, as shown in Figure \ref{fig:dists}, generally agree with the 50\% and 30\% targets, the overall mix of signal and background events are different.  This arises from the differing signal and background energy distributions near the selection threshold.  Overall, the HESE Gold stream has the highest average signal purity of $\sim 57\%$.

In addition to overall rates for each alert type, we also calculate the false alarm rate (FAR) on an event-by-event basis. The FAR for a given alert gives the annual rate of background events in IceCube with a direction and energy similar to the issued alert, and is derived from $N_{\rm{background}}(E,\delta)$ used in the signalness calculation. 


\begin{deluxetable}{c c c c c}[ht!]
\tabletypesize{\small}
\tablecaption{The number of expected signal and background events, and the total observed events for each alert stream in $\sim$9.6 years of the catalog live time. The expected number of events are calculated for the best-fit diffuse muon neutrino flux~\citep{IceCube:2021uhz} with a spectral index of $2.37$.}
\label{tab:rates}

\tablehead{
\colhead{Event Type} & \colhead{Expected Signal} & \colhead{Expected Background} & \colhead{Total Expected} & \colhead{Total Observed}
}
\startdata
 GFU Gold & 54.3 & 47 & 101.3 & 72 \\
 GFU Bronze & 40.2 & 138 & 178.2 & 164 \\
 HESE Gold & 5.3 & 4 & 9.3 & 9\\
 HESE Bronze & 1.6 & 9 & 10.6 & 8 \\
 EHE Gold & 3.9 & 19 & 22.9 & 22\\
\enddata
\end{deluxetable}
\section{Alert Processing and Follow-ups}\label{sec:alerts}
Due to computational limitations at the South Pole experimental site, and the need for promptly issuing an alert, the online reconstruction cannot utilize complex, computationally-intensive algorithms. Once all the event information has been transmitted to the North, a more refined set of follow-up reconstructions based on \cite{IceCube:2013dkx} begin on a computing cluster. The follow-up reconstruction consists of a maximum-likelihood based scan of the entire sky to search for an event direction consistent with the signals registered by the DOMs. We bin the sky into two-dimensional grids of increasing resolution in steps following the HEALPix pixelization scheme \citep{Gorski:2004by}. Each pixel defines a potential event direction in right ascension and declination. At each step, in each pixel, we fix the event direction and compute the likelihood of the best-fit deposited energy and the neutrino interaction position. Repeating the procedure over all pixels yields a likelihood map of the sky. The pixel corresponding to the maximum likelihood defines the best-fit direction of the neutrino event.  The scans are first performed on a coarse grid with NSIDE = 8, corresponding to a mean pixel spacing of $7.3^\circ$. The best-fit pixels are selected for the next steps with finer scans with NSIDES 64 (pixel size $0.9^\circ$) and 1024 (pixel size $0.06^\circ$). Each sky scan takes between one to three hours, yielding an improved angular reconstruction over the initial alert. In order to ensure that the reconstruction converges to a global minimum, at each step, we test several positional variations iteratively by using the result of a regular fit as a seed for the next iteration. If there are multiple local minima, the repeated iterations ensure that at least one of the results is a global minimum.

The angular uncertainty contours at 50\% and 90\% confidence level are extracted using predefined values of change in log-likelihood based on simulated neutrino events to ensure the required coverage \citep{txs2}. In order to cover potential systematic errors from detector uncertainties (such as glacial ice optical parameters), the detector systematic parameters were varied within expected errors during the simulation of the neutrino events used to determine the contour levels. As described in \cite{txs2}, we simulate an ensemble of events with similar energy deposition and position in the detector as IC160427A \citep{2019A&A...626A.117P}, varying the ice model parameters \citep{tc-2022-174} in each simulation. The simulated events are reconstructed, and the log-likelihoods of their reconstructed directions are compared to the log-likelihoods of their true directions. This procedure yields a distribution of change in log-likelihood which folds in the systematic uncertainties, and is used to extract the error contours.  We note that the calibration of the change in log-likelihood is especially sensitive to the optical properties of the ice, an area of intense study within IceCube \citep{IceCube:2021uwf}.  While these results represent our current modeling, updated parameters and reconstructions for alert candidates will be released as catalog updates as they become available. In real-time operations, the results from the follow-up scan are disseminated via a GCN circular and a revised GCN notice. In particular, the notice includes the circularized error region based on the follow-up reconstruction. An example of the revised  GCN notice for the event IC190730A can be seen at this URL \url{https://gcn.gsfc.nasa.gov/gcn3/25225.gcn3}.

\begin{figure}[h]
\includegraphics[width=1\textwidth]{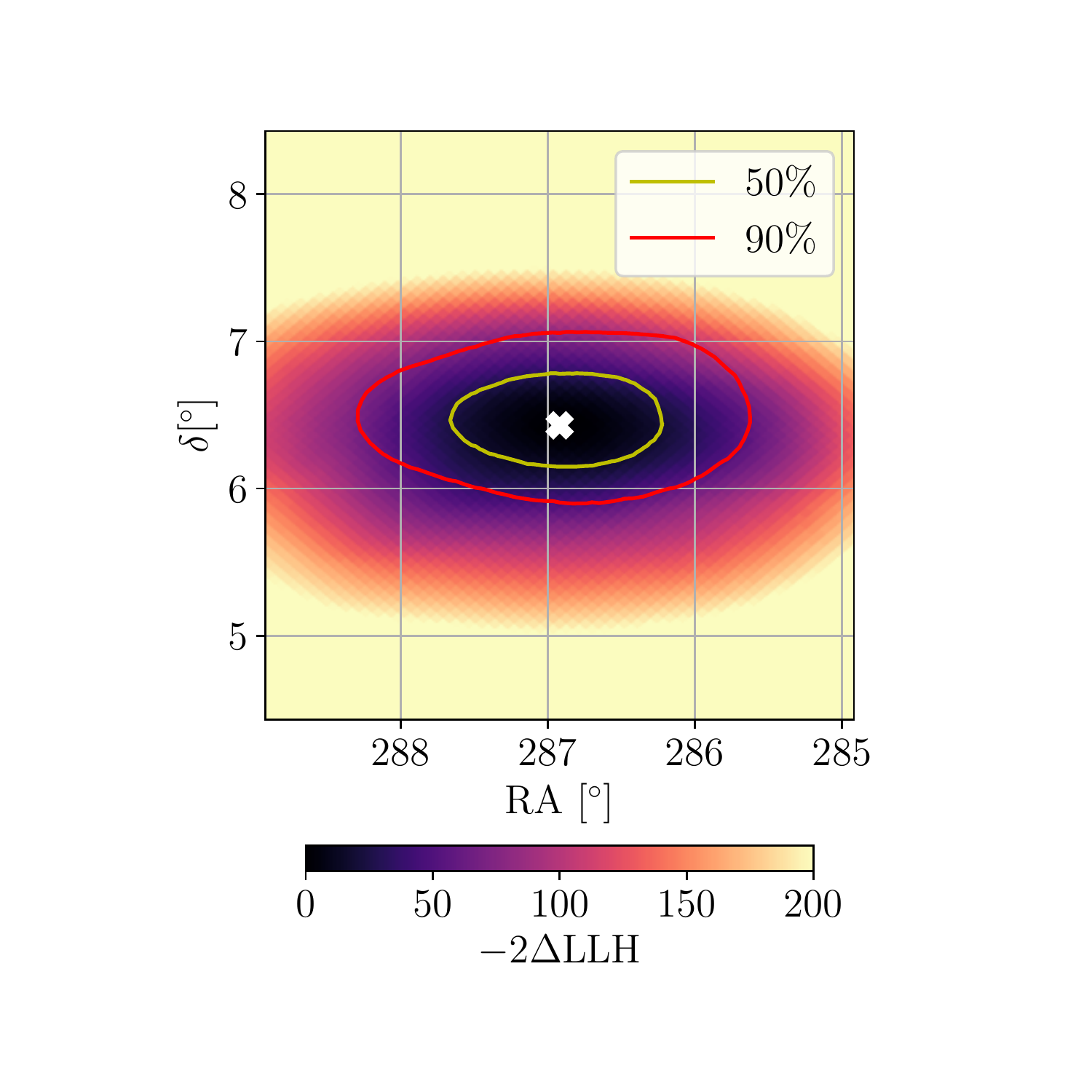}
\caption{The result of the sky scan shown in likelihood space for IC150119.
The best-fit position of the event is noted by the white cross in the center where
the likelihood is maximized. The yellow and red contours denote the uncertainty
in the location as the 50\% and 90\% confidence level changes in the likelihood (see section \ref{sec:alerts}). A rectangle that contains the 90\% error contour is used to report the +/- errors on the alert location (see Table \ref{tab:sources}).}
\label{fig:scan}
\end{figure}

For the compilation of the catalog, the same sky scan is performed on the selected alerts from the archival IceCube data on a commercial cloud computing service. Figure \ref{fig:scan} illustrates the result of one such scan, for an example alert from the catalog, IC150119. In addition, we also apply a convolutional neural network (CNN) based classifier to better distinguish the morphology of each event \citep{2019ICRC...36..937K}. Each event is assigned a score between 0 and 1 for how well it fits the following four hypotheses: cascade, skimming event (primary vertex outside the detector and no energy deposited within), starting track (interaction vertex inside the detector volume) or stopping track (track length of less than $\sim 1500$ m).  In this work, we provide the complete likelihood maps and the uncertainty contours, as described above,  for all the alerts in the accompanying data release.

\subsection{IceTop Veto}
Recently, 21 Oct 2022, we introduced an additional veto mechanism to reject atmospheric muons that may pass the alert selection criteria. This veto makes use of the IceTop surface array to search for cosmic-ray-induced showers accompanying a track-like event in the in-ice detector \citep{Amin_2021}. This is particularly useful in the case of down-going air showers inclined at an angle, typically below $82^\circ$ with respect to the zenith, that pass through IceCube and where the reconstructed track is not fully contained in the IceTop detector footprint. The IceTop veto criteria look for a threshold number of coincident pulses in IceTop tanks during a 1 $\mu$s time-window \citep{Amin_2021}. Since the criteria were developed after the compilation of the catalog, we do not discard the vetoed alerts but mark them as such. The probability that the veto algorithm incorrectly rejects a true astrophysical neutrino event is $\sim 10^{-4}$.

\section{Catalog Properties}\label{sec:cat}
\begin{figure}[h]
\includegraphics[width=1\textwidth, trim=0cm 0cm 0cm 0cm, clip=true]{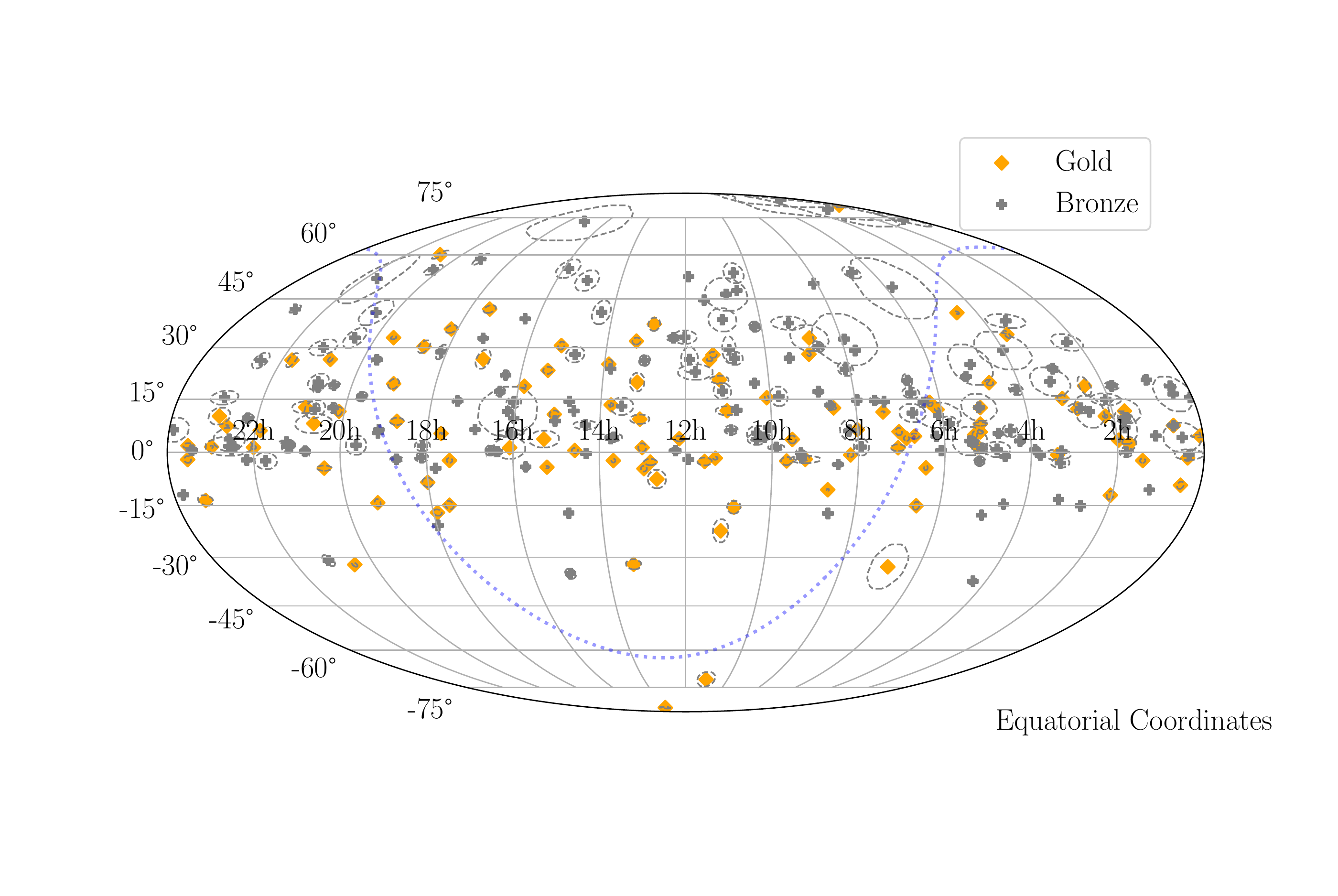}
\caption{The all-sky distribution of the alerts in the catalog in equatorial coordinates. The orange diamonds show the Gold alerts. The gray crosses show the Bronze alerts. The 90\% uncertainty contours at the location of each alert are shown by the dashed ellipses.}
\label{fig:map}
\end{figure}

\begin{figure}[tbp!]
\includegraphics[width=0.33\textwidth, trim=0cm 0cm 0cm 1.5cm, clip=true]{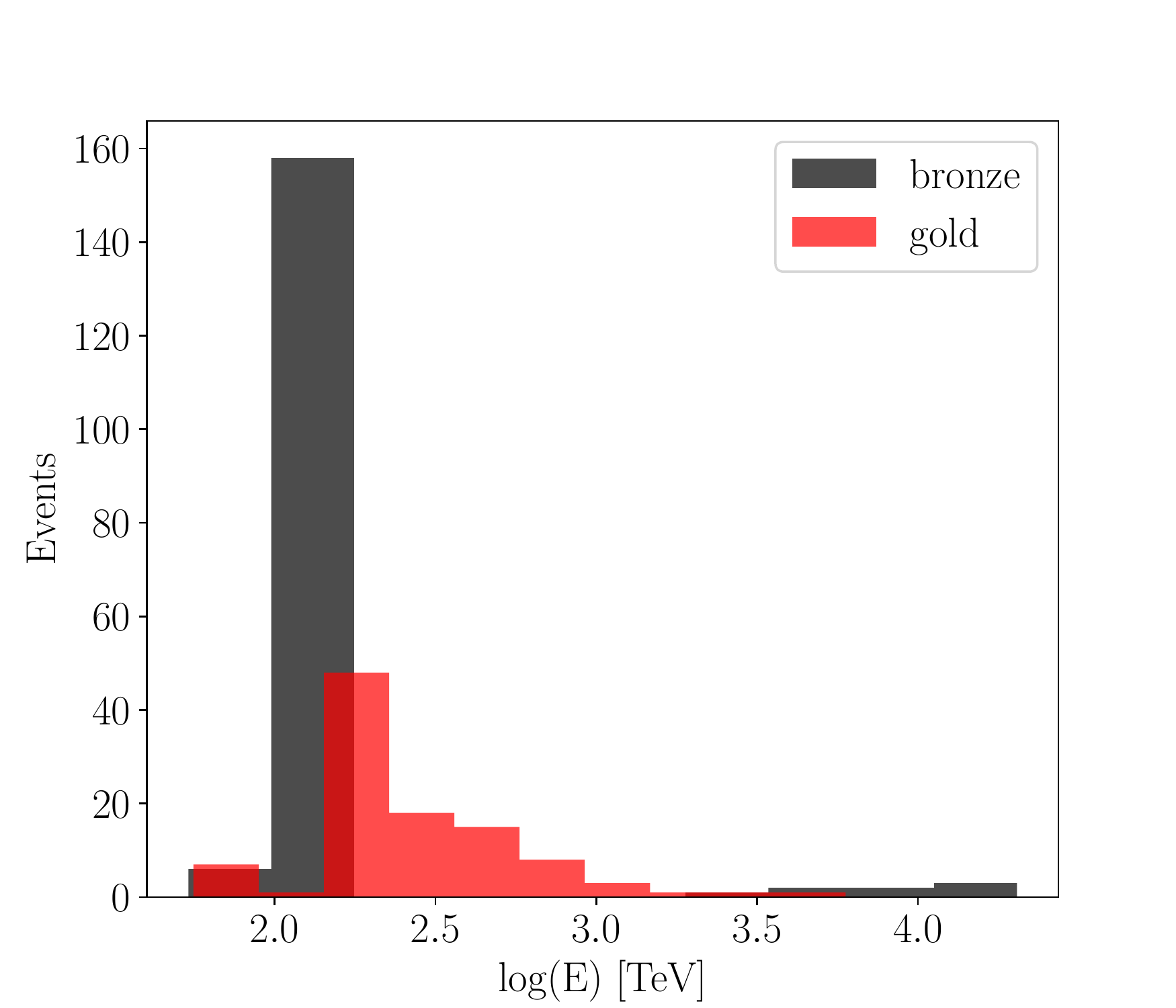}\includegraphics[width=0.33\textwidth, trim=0cm 0cm 0cm 1.5cm, clip=true]{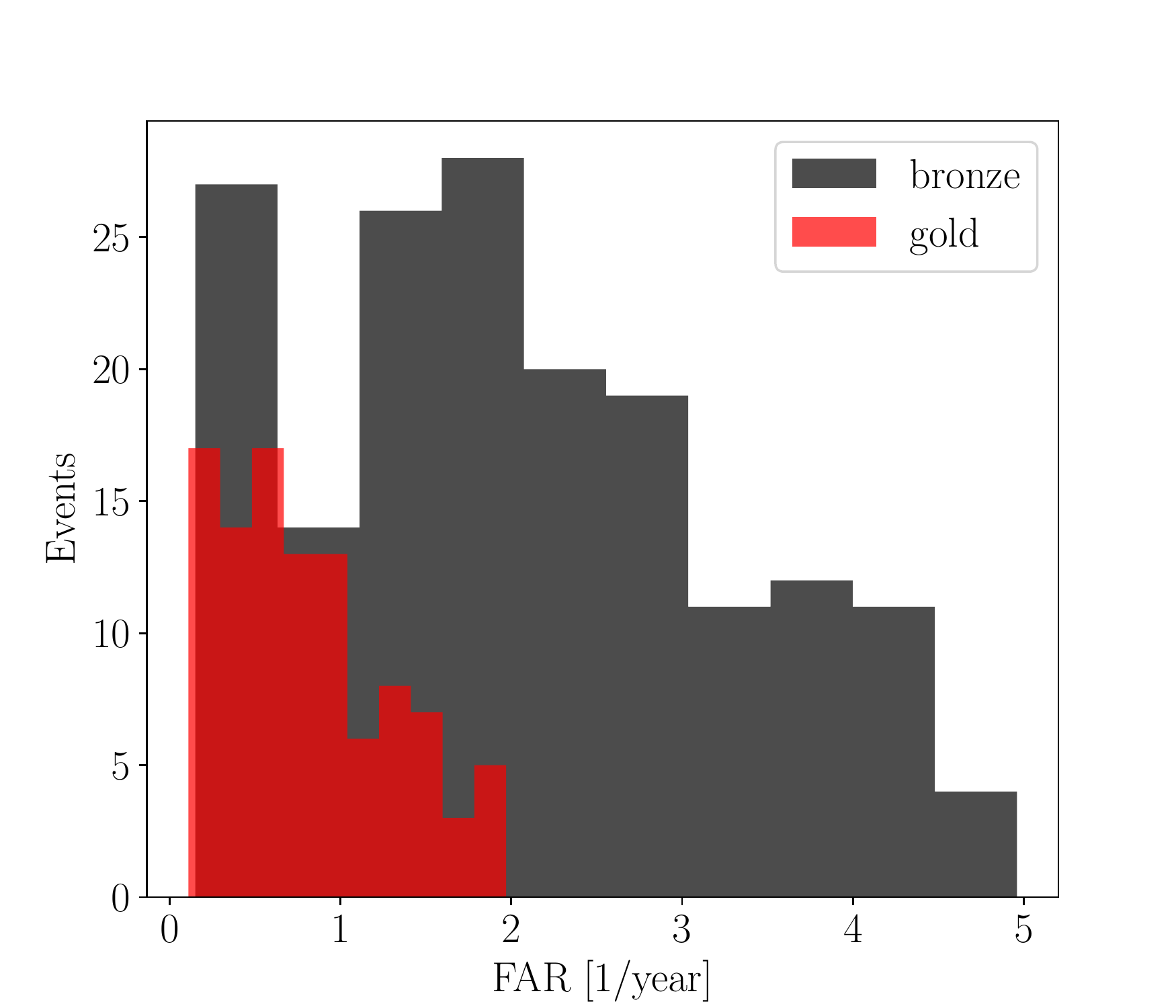} \includegraphics[width=0.33\textwidth, trim=0cm 0cm 0cm 1.5cm, clip=true]{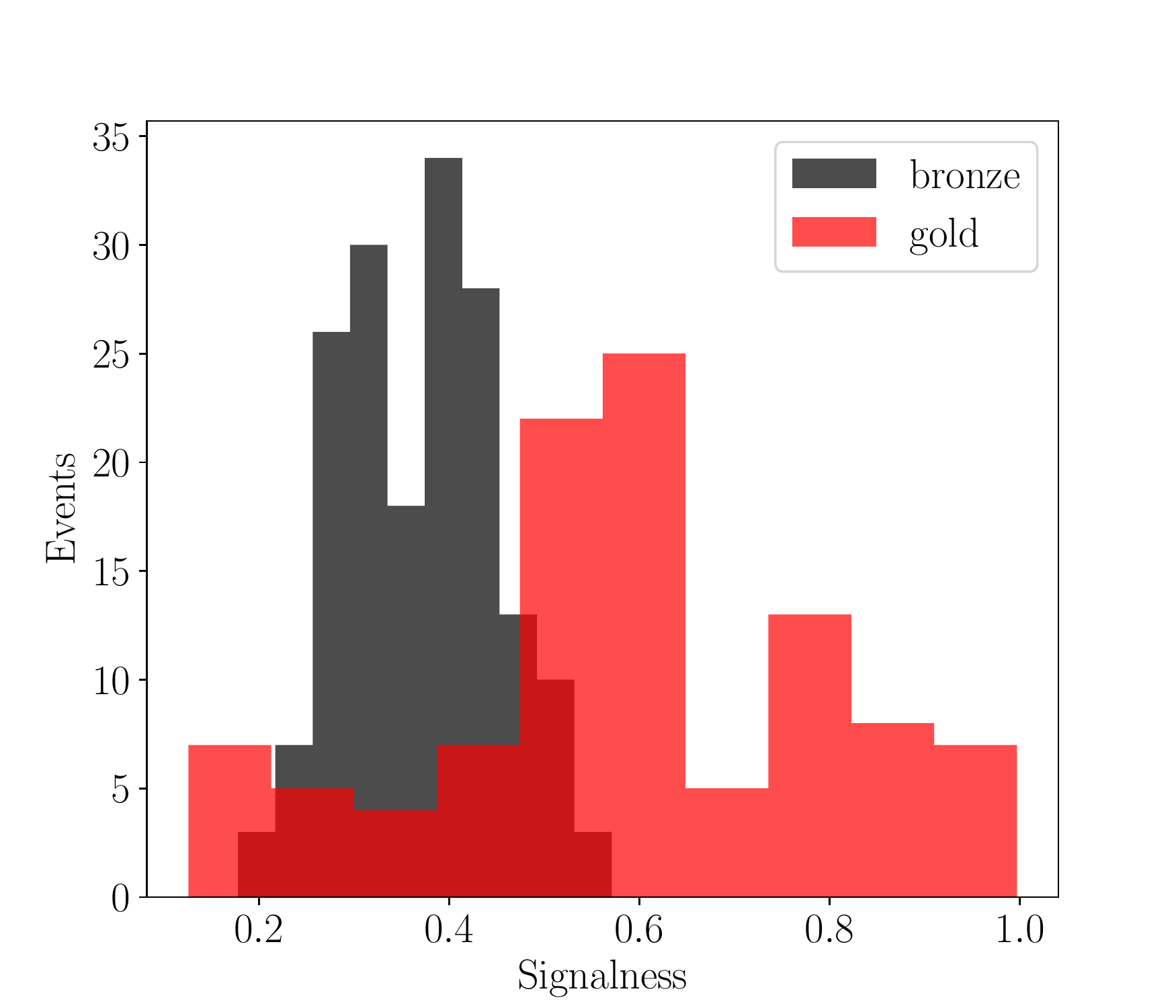}
\caption{\textit{Left}: The logarithmic distribution of likely neutrino energies (assuming a spectral index of $2.19$)  in TeV for the catalog.
\textit{Center}: The distribution of the false alarm rate of the alerts in the catalog.
\textit{Right}: The distribution of signalness of the alerts in the catalog. The red (grey)
histogram shows the Gold (Bronze) alerts.}
\label{fig:dists}

\end{figure}

We compile the neutrino alert catalog by applying the above-mentioned procedures of event selection followed by likelihood scans on IceCube data going back to May 2011. A total of 275 events pass the alert criteria through the end of 2020, including alerts issued in real-time after the updated system was activated in June 2019. Figure \ref{fig:map} shows the location of all the alerts on a sky map in equatorial coordinates. Figure \ref{fig:map2} shows the distribution broken down by alert type. The breakdown of the number of alerts by stream and selection is shown in Table \ref{tab:rates}. Figure \ref{fig:dists} shows the distributions of energies, false alarm rates, and signalness parameters for all the alerts. Table \ref{tab:sources} shows all alerts with their best-fit directions and 90\% uncertainties (J2000 coordinates), energy (assuming a spectral index of 2.19) and signalness information. The probable neutrino energy for each event is calculated from the observed muon energy \citep{2013NIMPA.703..190A}. Figure \ref{fig:energy_mc} illustrates the spread of true neutrino energies that contribute to a given observed muon energy. The uncertainties on alert directions reported in Table \ref{tab:sources} are obtained from the rectangular region bounding the error contours. After the construction of the catalog, we also checked data from IceTop for signatures of cosmic-ray activity that is temporally correlated with each of the alerts. Eight alerts were vetoed by IceTop data as described above. Such alerts are likely to be caused by atmospheric background and are marked with an asterisk (*) in Table \ref{tab:sources}, but as these veto criteria were added at a later time, these events were not removed from the catalog.  These vetoed events are likely not of astrophysical origin, and future real-time alerts will not be issued for events that fail these veto criteria.

The neutrino event selection used in this catalog is designed to select astrophysical event candidates that are likely to provide well-reconstructed directions on the sky.  However, this is not the only astrophysical event selection to have been used in IceCube, and several historical catalogs of astrophysical neutrino candidates have been previously released~\citep{txs2,IceCube:2014stg,IceCube:2016umi,IceCube:2021uhz}.  
While several events contained in these previous catalogs are included here, several do not meet the selection criteria used for this real-time alert selection.  This does not imply that these events are not potential astrophysical neutrino candidates, but rather that automated event selections can not supply sufficient information to issue alerts automatically. The information included here for these events also represents our updated understanding of these candidates, using the latest calibration, glacial ice modeling, and reconstruction algorithms.  

The complete catalog is provided in electronic format on \href{https://dataverse.harvard.edu/dataset.xhtml?persistentId=doi:10.7910/DVN/SCRUCD}{dataverse}\footnote{https://dataverse.harvard.edu/dataset.xhtml?persistentId=doi:10.7910/DVN/SCRUCD} with columns in addition to the ones in table \ref{tab:sources} for each event as follows: \texttt{I3TYPE} (event selection type), \texttt{FAR} per year, scores for each CNN classification for event topology, and a \texttt{CR\_VETO} flag to mark significant temporally coincident cosmic-ray shower activity with a given alert.    

\section{Search for Correlations with Potential Candidates for Association}\label{sec:assc}
Once an IceCube alert is issued, telescopes can begin follow-up observations around the best-fit location of the neutrino event for potential electromagnetic counterparts. Sources lying within the angular uncertainty contours can be probed on different time-scales for transient activity to obtain clues as to the origin of the neutrino. Using a variety of data samples, IceCube continues to conduct searches for long-term neutrino emission and for correlations of neutrino data with known astrophysical objects across different wavelengths \citep{icecube_7year,ngc,2019ApJ...886...12A,IceCube:2021waz}. For the 275 neutrino events in this catalog, we are performing several follow-up analyses that are the subject of on-going and future publications \citep{IceCubeCollaboration:2022fxl,IceCube:2023htm}. In this work, we report the results of a time-independent search for spatial correlations between the best-fit positions of the alerts and of sources from five catalogs of gamma-ray and X-ray sources. 

We use the following catalogs: the 4FGL-DR2 \citep{Fermi-LAT:2019yla} and 3FHL \citep{Fermi-LAT:2017sxy} catalogs from  Fermi-LAT, 3HWC catalog \citep{HAWC:2020hrt} from the HAWC observatory, TeVCat \citep{2008ICRC....3.1341W}, and the \textit{Swift}-BAT catalog of hard X-ray sources \citep{2018ApJS..235....4O}. We note that these catalogs are not completely independent and some of the sources are present in multiple catalogs. For each of the 275 alerts, using the aforementioned catalogs we search for sources that lie within the 90\% uncertainty contour of the alert's reconstructed direction. For all sources found within the error contour of a given alert, we calculate their angular distance from the best-fit location of the alert. The closest source and its distance from the best-fit location of the alert are reported in Table \ref{tab:sources}. We find that 139 neutrino alerts have no source from any of the above catalogs in the uncertainty region. For each of the five catalogs, we also determine the total number of alerts that are spatially coincident with at least one source in the catalog. We also determine how many such coincidences are expected due to chance by randomizing the alert directions in right ascension 1000 times and looking at the number of coincidences after each randomization. The sensitivity of IceCube is approximately uniform as a function of right ascension. The randomization allows the production of simulated data with the characteristics of the null hypothesis (no correlation with the catalogs). For each catalog, we find that the number of coincidences is consistent with the median expectation due to chance. The number of observed correlations and the median number expected due to chance for each catalog are shown in Table \ref{tab:coinc}.


\begin{deluxetable}{c c c}[ht!]
\tabletypesize{\small}
\tablecaption{The number of alerts with a particular catalog source located within the error contours, and the number of such observations expected due to chance.}
\label{tab:coinc}

\tablehead{
\colhead{Catalog} & \colhead{Observed Coincidences} & \colhead{Expected Coincidences}
}
\startdata
4FGL & 119 & 140 \\
3FHL & 67 & 77\\
3HWC & 8 & 6 \\
TeVCat & 12 & 16\\
BAT & 66 & 73
\enddata
\end{deluxetable}

Five Fermi-LAT sources -- 4FGL J0914.1-0202, 4FGL J1019.7+0511, 4FGL J2226.6+0210, 4FGL J2227.9+0036  and 4FGL J0244.7+1316 --  and one Swift-BAT source, SWIFT J2235.7+013, appear to be spatially correlated with more than one alert, and are considered as repeated candidates for association. The correlations however are not unique as there are often multiple sources located within an error contour. Moreover, such repeated associations are not uncommon in randomized right ascension data sets. For 4FGL alone, we observe on average four candidates for repeated associations in 1000 simulations. We emphasize that spatial correlations between neutrino alerts and sources from other catalogs are not evidence for definitive association as the observed number of correlations is consistent with accidental correlations, as shown above. However, we encourage dedicated follow-up studies using the light-curves of the closest sources to each alert identified in this study.        
\section{Summary and Conclusion}\label{sec:conclusion}
Neutrinos play an extremely important role in the era of multimessenger astronomy, serving as our windows into the  complex physics underlying cosmic-ray accelerators. To this end, IceCube has an active program dedicated to immediately alerting the community of a potential astrophysical neutrino detection. The program began in 2016, with significant improvements following in 2019 that are described in this work. Here, we provide a catalog, the IceCube Event Catalog of Alert Tracks (ICECAT-1), of 275 track-like neutrino events that retroactively pass the alert criteria from 2011 to 2020. The event information for each alert is available to the public in the form of FITS \citep{2010A&A...524A..42P} files that include the complete likelihood profiles providing an accurate grasp on the spatial uncertainty. This catalog, as well as updates with additional alerts, can be found at  \url{https://doi.org/10.7910/DVN/SCRUCD}.  This format will also be introduced for future IceCube alerts in addition to the traditional GCN notice format mode of distribution. We have also explored the correlation of IceCube alerts with sources from very-high-energy gamma-ray and X-ray catalogs and find them consistent with chance expectation. Future IceCube analyses, will more systematically explore the correlation of these alerts with blazars, as well as with other IceCube data on long and short time-scales \citep{IceCube:2020mzw,IceCube:2023htm}. Several observatories have dedicated programs to searching for electromagnetic counterparts of the IceCube real-time alerts leading to the identification of potential sites of cosmic-ray acceleration \cite{10.1093/mnras/stac2261, Plavin:2020emb,Dzhappuev:2020bkh,Stein:2020xhk}. Multi-wavelength follow-up observations will also benefit from the information provided about IceCube alerts in this catalog. We also note that a revised reconstruction framework is in the works that will improve the angular errors for alerts in the future.


\startlongtable
  \begin{deluxetable*}{l c c c c c l}
    \tabletypesize{\small}
    \tablecaption{Alert events in the catalog along with their time, positions, energy, signalness, and the closest source within the alert error contours from the spatial correlation search. The distance to the coincident sources is shown in parentheses with each source name. Events marked with an asterisk (*) also triggered IceTop and are likely due to cosmic-ray showers. The errors on RA and Dec correspond to the 90\% uncertainty likelihood contours  (see text).
      \label{tab:sources}}
    \tablehead{
      \colhead{Alert} & \colhead{MJD} & \colhead{RA} & \colhead{Dec} & \colhead{Energy} & \colhead{Signalness} & \colhead{Nearest Source ($^\circ$)}  \\
      \colhead{} &  \colhead{} & \colhead{[$^\circ$]} & \colhead{[$^\circ$]} & \colhead{[TeV]} & \colhead{} & \colhead{}
    }
    \startdata
      IC110514A & 55695.064 & $138.47^{+6.68}_{-3.78}$ & $-1.94^{+0.97}_{-1.12}$ & 187 & $0.51$ & 4FGL J0914.1-0202 (0.12)   \\
IC110610A & 55722.426 & $272.55^{+1.67}_{-2.42}$ & $35.64^{+1.30}_{-1.05}$ & 294 & $0.75$ & 4FGL J1808.8+3522 (0.37)   \\
IC110616A & 55728.730 & $71.15^{+1.41}_{-2.07}$ & $5.38^{+0.79}_{-0.90}$ & 109 & $0.26$ & ... (...)   \\
IC110714A & 55756.113 & $68.20^{+0.31}_{-1.10}$ & $40.67^{+0.44}_{-0.44}$ & 72 & $0.78$ & ... (...)   \\
IC110726A & 55768.511 & $151.08^{+1.19}_{-1.71}$ & $6.99^{+0.98}_{-0.83}$ & 160 & $0.40$ & ... (...)   \\
IC110807A & 55780.980 & $336.80^{+1.36}_{-1.98}$ & $1.53^{+0.93}_{-0.78}$ & 108 & $0.27$ & 4FGL J2226.6+0210 (0.65)   \\
IC110818A & 55791.689 & $332.45^{+0.97}_{-1.23}$ & $-2.09^{+0.93}_{-0.90}$ & 123 & $0.34$ & ... (...)   \\
IC110902A & 55806.092 & $9.76^{+2.86}_{-1.32}$ & $7.59^{+0.87}_{-0.86}$ & 243 & $0.61$ & ... (...)   \\
IC110907A & 55811.795 & $196.08^{+3.91}_{-2.68}$ & $9.40^{+1.55}_{-1.06}$ & 186 & $0.51$ & 4FGL J1301.6+0834 (1.06)   \\
IC110929A & 55833.260 & $121.45^{+1.34}_{-1.29}$ & $50.04^{+0.24}_{-0.15}$ & 158 & $0.52$ & ... (...)   \\
IC110930A & 55834.445 & $267.01^{+1.19}_{-1.14}$ & $-4.44^{+0.60}_{-0.79}$ & 160 & $0.43$ & ... (...)   \\
IC111012A & 55846.867 & $172.13^{+1.40}_{-1.39}$ & $44.70^{+0.79}_{-0.45}$ & 115 & $0.43$ & ... (...)   \\
IC111120A & 55885.961 & $26.06^{+1.89}_{-3.16}$ & $9.82^{+1.40}_{-1.36}$ & 159 & $0.42$ & ... (...)   \\
IC111120B* & 55885.973 & $356.84^{+0.53}_{-0.62}$ & $-11.99^{+0.23}_{-0.27}$ & 4969 & $0.29$ & ... (...)   \\
IC111208A & 55903.719 & $165.19^{+7.03}_{-4.13}$ & $38.49^{+3.67}_{-3.49}$ & 123 & $0.45$ & 4FGL J1101.5+3904 (0.6)   \\
IC111209A & 55904.457 & $99.98^{+1.19}_{-2.02}$ & $20.42^{+1.60}_{-2.02}$ & 108 & $0.34$ & 3HWC J0633+191 (1.95)   \\
IC111213A & 55908.398 & $247.85^{+1.71}_{-1.58}$ & $0.56^{+1.46}_{-1.42}$ & 164 & $0.41$ & 4FGL J1638.0+0042 (1.67)   \\
IC111216A & 55911.277 & $36.74^{+1.80}_{-2.24}$ & $18.88^{+2.46}_{-2.82}$ & 891 & $0.95$ &    SWIFT J0225.0+18 (0.46)   \\
IC111218A & 55913.335 & $26.85^{+3.69}_{-4.66}$ & $7.03^{+4.04}_{-5.20}$ & 157 & $0.40$ & 3FHL J0151.0+0539 (1.64)   \\
IC120301A & 55987.807 & $237.96^{+0.53}_{-0.62}$ & $18.76^{+0.47}_{-0.51}$ & 433 & $0.82$ & ... (...)   \\
IC120426A & 56043.415 & $183.56^{+2.15}_{-2.02}$ & $0.52^{+0.86}_{-0.71}$ & 109 & $0.27$ & ... (...)   \\
IC120501A & 56048.570 & $165.37^{+7.01}_{-5.36}$ & $-71.51^{+3.53}_{-2.68}$ & 85 & $0.46$ & ... (...)   \\
IC120515A & 56062.959 & $198.94^{+1.71}_{-1.41}$ & $32.00^{+0.97}_{-1.09}$ & 194 & $0.61$ & ... (...)   \\
IC120523A & 56070.574 & $171.08^{+0.66}_{-1.41}$ & $26.44^{+0.46}_{-0.37}$ & 213 & $0.53$ & ... (...)   \\
IC120523A & 56070.639 & $343.78^{+4.92}_{-4.48}$ & $15.48^{+2.38}_{-1.54}$ & 168 & $0.49$ & 4FGL J2253.9+1609 (0.72)   \\
IC120529A & 56076.543 & $176.48^{+6.64}_{-5.93}$ & $22.87^{+2.70}_{-1.77}$ & 126 & $0.42$ &    SWIFT J1141.3+21 (1.45)   \\
IC120601A & 56079.306 & $119.31^{+2.02}_{-0.92}$ & $14.79^{+0.62}_{-0.73}$ & 137 & $0.40$ & ... (...)   \\
IC120605A & 56083.655 & $152.58^{+1.89}_{-2.42}$ & $36.38^{+1.54}_{-1.47}$ & 107 & $0.39$ & 4FGL J1011.6+3600 (0.45)   \\
IC120611A* & 56089.364 & $39.95^{+0.26}_{-0.26}$ & $-15.09^{+0.19}_{-0.31}$ & 9220 & $0.24$ & ... (...)   \\
IC120807A & 56146.207 & $330.07^{+0.83}_{-0.83}$ & $1.42^{+0.60}_{-0.45}$ & 373 & $0.74$ & ... (...)   \\
IC120916A & 56186.305 & $182.24^{+1.36}_{-1.71}$ & $3.88^{+0.67}_{-0.82}$ & 174 & $0.44$ & 4FGL J1204.8+0407 (1.06)   \\
IC120922A & 56192.549 & $70.62^{+1.49}_{-1.27}$ & $19.79^{+0.91}_{-0.71}$ & 143 & $0.43$ & ... (...)   \\
IC121011A & 56211.771 & $205.14^{+0.66}_{-0.70}$ & $-2.28^{+0.52}_{-0.56}$ & 481 & $0.84$ & ... (...)   \\
IC121026A & 56226.599 & $169.80^{+1.32}_{-1.41}$ & $27.91^{+0.85}_{-0.88}$ & 961 & $0.93$ & ... (...)   \\
IC121103A & 56234.508 & $123.18^{+0.92}_{-0.97}$ & $6.05^{+0.64}_{-0.56}$ & 112 & $0.28$ & ... (...)   \\
IC121115A & 56246.330 & $225.70^{+1.01}_{-1.19}$ & $8.88^{+0.94}_{-0.95}$ & 116 & $0.32$ & ... (...)   \\
IC130125A & 56317.266 & $7.67^{+6.46}_{-5.92}$ & $74.14^{+3.36}_{-2.82}$ & 165 & $0.53$ & 4FGL J0028.1+7505 (0.97)   \\
IC130125A & 56317.659 & $280.46^{+1.89}_{-2.33}$ & $-1.90^{+0.93}_{-0.82}$ & 114 & $0.31$ & 4FGL J1847.2-0141 (1.37)   \\
IC130127A & 56319.280 & $352.97^{+1.32}_{-1.01}$ & $-1.98^{+0.97}_{-0.90}$ & 235 & $0.61$ & 4FGL J2333.4-0133 (0.58)   \\
IC130208A* & 56331.121 & $48.38^{+0.31}_{-0.31}$ & $-13.32^{+0.23}_{-0.23}$ & 2268 & $0.32$ & ... (...)   \\
IC130316A & 56367.736 & $303.41^{+2.98}_{-3.55}$ & $54.68^{+1.77}_{-1.59}$ & 105 & $0.38$ &    SWIFT J2006.5+56 (1.93)   \\
IC130318A & 56369.285 & $13.45^{+0.79}_{-0.88}$ & $20.62^{+1.44}_{-0.99}$ & 106 & $0.33$ & ... (...)   \\
IC130408A & 56390.189 & $167.83^{+2.64}_{-3.96}$ & $20.66^{+1.28}_{-0.99}$ & 65 & $0.53$ &    SWIFT J1114.3+20 (0.71)   \\
IC130408B & 56390.758 & $7.38^{+4.88}_{-8.04}$ & $4.22^{+4.73}_{-3.58}$ & 163 & $0.40$ & 4FGL J0030.4+0451 (0.68)   \\
IC130409A & 56391.982 & $163.56^{+2.68}_{-2.50}$ & $29.44^{+4.38}_{-3.46}$ & 115 & $0.41$ & GB6 J1058+2817 (1.48)   \\
IC130508A & 56420.641 & $337.76^{+3.21}_{-2.02}$ & $26.24^{+2.69}_{-1.90}$ & 140 & $0.45$ &    SWIFT J2237.0+25 (1.35)   \\
IC130509A & 56421.186 & $317.50^{+1.76}_{-1.85}$ & $2.09^{+1.19}_{-1.34}$ & 105 & $0.25$ & 4FGL J2104.7+0108 (1.63)   \\
IC130519A & 56431.483 & $45.35^{+3.12}_{-1.49}$ & $23.85^{+1.15}_{-0.89}$ & 110 & $0.36$ & ... (...)   \\
IC130531A & 56443.557 & $164.18^{+2.42}_{-2.15}$ & $6.32^{+1.32}_{-1.27}$ & 143 & $0.35$ & ... (...)   \\
IC130627A & 56470.110 & $93.74^{+1.01}_{-1.14}$ & $14.17^{+1.23}_{-1.04}$ & 851 & $0.94$ & 4FGL J0615.9+1416 (0.26)   \\
IC130627B & 56470.426 & $155.35^{+3.87}_{-2.37}$ & $3.73^{+1.72}_{-1.42}$ & 122 & $0.31$ & ... (...)   \\
IC130711A & 56484.530 & $77.87^{+2.59}_{-1.19}$ & $-2.43^{+1.34}_{-1.12}$ & 165 & $0.43$ & 4FGL J0515.5-0125 (1.44)   \\
IC130731A & 56504.072 & $122.87^{+2.29}_{-4.35}$ & $6.32^{+3.24}_{-2.40}$ & 122 & $0.32$ & 4FGL J0812.5+0711 (0.92)   \\
IC130801A & 56505.256 & $214.98^{+4.35}_{-4.00}$ & $7.75^{+1.24}_{-1.20}$ & 110 & $0.28$ &    SWIFT J1419.1+07 (0.26)   \\
IC130804A & 56508.815 & $129.02^{+1.14}_{-1.54}$ & $13.36^{+1.08}_{-1.68}$ & 113 & $0.33$ & ... (...)   \\
IC130808A & 56512.340 & $26.59^{+1.14}_{-1.23}$ & $9.22^{+0.91}_{-0.87}$ & 111 & $0.29$ & ... (...)   \\
IC130822A & 56526.409 & $91.32^{+1.19}_{-1.19}$ & $0.56^{+0.78}_{-0.63}$ & 115 & $0.30$ & 3FHL J0604.9+0000 (0.56)   \\
IC130907A* & 56542.793 & $130.17^{+0.48}_{-0.31}$ & $-10.54^{+0.27}_{-0.30}$ & 890 & $0.32$ & ... (...)   \\
IC131014A & 56579.909 & $32.92^{+0.88}_{-0.70}$ & $10.28^{+0.42}_{-0.57}$ & 293 & $0.67$ & ... (...)   \\
IC131023A & 56588.559 & $301.90^{+1.01}_{-1.05}$ & $11.61^{+1.14}_{-1.29}$ & 211 & $0.59$ & ... (...)   \\
IC131108A & 56604.553 & $342.73^{+1.54}_{-1.58}$ & $41.81^{+1.40}_{-0.89}$ & 153 & $0.50$ & ... (...)   \\
IC131112A & 56608.031 & $129.24^{+0.26}_{-0.26}$ & $-17.27^{+0.16}_{-0.16}$ & 7006 & $0.27$ & ... (...)   \\
IC131124A & 56620.145 & $285.16^{+2.20}_{-1.54}$ & $19.47^{+1.43}_{-1.46}$ & 180 & $0.55$ & ... (...)   \\
IC131204A & 56630.470 & $288.98^{+1.10}_{-0.83}$ & $-14.21^{+0.77}_{-1.31}$ & 259 & $0.20$ & 4FGL J1916.7-1516 (1.08)   \\
IC140101A & 56658.404 & $192.26^{+2.07}_{-2.37}$ & $-2.69^{+1.01}_{-0.71}$ & 200 & $0.56$ & 4FGL J1251.3-0201 (0.88)   \\
IC140103A & 56660.886 & $37.90^{+25.61}_{-27.30}$ & $78.97^{+5.86}_{-9.97}$ & 125 & $0.42$ & 4FGL J0226.9+7744 (1.25)   \\
IC140108A & 56665.308 & $344.66^{+0.53}_{-0.48}$ & $1.57^{+0.37}_{-0.34}$ & 214 & $0.69$ & ... (...)   \\
IC140109A & 56666.503 & $293.12^{+0.79}_{-1.19}$ & $33.02^{+0.45}_{-0.53}$ & 924 & $0.93$ &    SWIFT J1933.9+32 (0.31)   \\
IC140114A & 56671.878 & $337.59^{+0.57}_{-0.92}$ & $0.71^{+0.97}_{-0.86}$ & 54 & $0.34$ & 4FGL J2227.9+0036 (0.61)   \\
IC140122A & 56679.147 & $138.82^{+3.52}_{-10.11}$ & $37.45^{+1.95}_{-2.09}$ & 131 & $0.46$ &    SWIFT J0920.1+37 (0.99)   \\
IC140122B & 56679.204 & $220.29^{+8.94}_{-8.37}$ & $-86.07^{+0.59}_{-0.64}$ & 374 & $0.82$ & ... (...)   \\
IC140203A & 56691.785 & $349.58^{+2.64}_{-2.55}$ & $-13.55^{+1.15}_{-1.73}$ & 685 & $0.13$ & ... (...)   \\
IC140213A & 56701.809 & $202.59^{+4.79}_{-3.21}$ & $13.06^{+2.31}_{-2.52}$ & 140 & $0.39$ & 4FGL J1326.1+1232 (1.14)   \\
IC140223A & 56711.920 & $118.83^{+11.87}_{-11.87}$ & $32.58^{+5.68}_{-9.83}$ & 119 & $0.43$ & 4FGL J0752.2+3313 (0.92)   \\
IC140307A & 56723.920 & $308.06^{+2.68}_{-4.61}$ & $32.93^{+2.71}_{-3.23}$ & 109 & $0.40$ & 4FGL J2028.3+3331 (1.01)   \\
IC140324A & 56740.089 & $225.70^{+5.67}_{-4.65}$ & $51.06^{+4.00}_{-2.87}$ & 109 & $0.40$ & 4FGL J1456.0+5051 (1.08)   \\
IC140410A & 56757.099 & $2.11^{+151.51}_{-58.92}$ & $81.22^{+8.00}_{-6.94}$ & 246 & $0.63$ &    SWIFT J0017.1+81 (0.5)   \\
IC140411A & 56758.567 & $146.95^{+3.12}_{-3.12}$ & $15.91^{+2.89}_{-2.62}$ & 156 & $0.45$ & 4FGL J0949.2+1749 (1.95)   \\
IC140420A & 56767.859 & $6.28^{+7.03}_{-5.89}$ & $16.57^{+4.77}_{-5.11}$ & 163 & $0.49$ & 4FGL J0023.9+1603 (0.58)   \\
IC140503A & 56780.957 & $162.30^{+6.91}_{-11.26}$ & $46.57^{+5.41}_{-5.11}$ & 109 & $0.40$ & 3FHL J1053.6+4930 (3.03)   \\
IC140603A & 56811.142 & $9.71^{+0.62}_{-0.88}$ & $7.56^{+0.53}_{-0.83}$ & 152 & $0.38$ & ... (...)   \\
IC140609A & 56817.636 & $106.26^{+2.68}_{-2.15}$ & $1.31^{+1.05}_{-0.86}$ & 459 & $0.81$ & ... (...)   \\
IC140611A & 56819.204 & $110.65^{+0.53}_{-0.62}$ & $11.45^{+0.19}_{-0.19}$ & 5960 & $1.00$ & ... (...)   \\
IC140704A & 56842.298 & $157.07^{+4.69}_{-4.64}$ & $53.62^{+3.35}_{-3.48}$ & 150 & $0.50$ &    SWIFT J1033.8+52 (1.07)   \\
IC140705A & 56843.669 & $25.88^{+1.85}_{-2.99}$ & $2.54^{+1.79}_{-1.75}$ & 212 & $0.56$ & 4FGL J0138.5+0300 (1.33)   \\
IC140707A & 56845.500 & $240.86^{+3.08}_{-2.07}$ & $14.17^{+1.54}_{-1.65}$ & 167 & $0.48$ & 4FGL J1606.2+1346 (0.8)   \\
IC140713A & 56851.557 & $0.79^{+1.14}_{-1.19}$ & $15.60^{+0.89}_{-0.66}$ & 134 & $0.39$ & ... (...)   \\
IC140721A & 56859.759 & $101.82^{+6.77}_{-6.86}$ & $-32.89^{+5.23}_{-8.08}$ & 157 & $0.56$ & 4FGL J0649.5-3139 (1.32)   \\
IC140820A & 56889.378 & $271.45^{+3.43}_{-1.80}$ & $1.87^{+1.42}_{-1.46}$ & 108 & $0.27$ & ... (...)   \\
IC140923A & 56923.721 & $169.72^{+0.70}_{-0.83}$ & $-1.60^{+0.52}_{-0.30}$ & 209 & $0.24$ & ... (...)   \\
IC140927A & 56927.161 & $50.89^{+3.91}_{-5.14}$ & $-0.63^{+1.49}_{-1.42}$ & 182 & $0.48$ & 3FHL J0323.6-0109 (0.54)   \\
IC141012A & 56942.751 & $63.85^{+2.24}_{-1.36}$ & $3.21^{+0.90}_{-1.08}$ & 173 & $0.44$ & 4FGL J0412.3+0239 (0.94)   \\
IC141110A & 56971.297 & $253.43^{+0.83}_{-1.10}$ & $6.43^{+0.71}_{-0.68}$ & 113 & $0.29$ & ... (...)   \\
IC141114A & 56975.257 & $221.48^{+4.53}_{-2.29}$ & $28.00^{+2.31}_{-2.30}$ & 110 & $0.38$ & 3FHL J1449.5+2745 (0.84)   \\
IC141208A & 56999.668 & $246.36^{+1.76}_{-1.89}$ & $17.23^{+1.29}_{-1.09}$ & 109 & $0.33$ & 4FGL J1626.4+1820 (1.13)   \\
IC141210A & 57001.848 & $318.12^{+2.33}_{-1.93}$ & $1.57^{+1.57}_{-1.72}$ & 154 & $0.37$ & ... (...)   \\
IC141221A & 57012.410 & $179.08^{+0.88}_{-1.10}$ & $-1.94^{+0.71}_{-0.82}$ & 134 & $0.35$ & ... (...)   \\
IC150102A & 57024.796 & $318.74^{+3.96}_{-1.27}$ & $2.91^{+0.34}_{-0.49}$ & 126 & $0.32$ & ... (...)   \\
IC150104A & 57026.399 & $272.11^{+1.71}_{-1.54}$ & $28.76^{+2.41}_{-1.86}$ & 133 & $0.45$ & 4FGL J1807.1+2822 (0.48)   \\
IC150118A & 57040.509 & $152.53^{+1.54}_{-2.72}$ & $4.33^{+0.71}_{-0.86}$ & 156 & $0.37$ & ... (...)   \\
IC150119A & 57041.369 & $286.92^{+1.36}_{-1.27}$ & $6.43^{+0.60}_{-0.53}$ & 140 & $0.35$ & 3HWC J1908+063 (0.14)   \\
IC150120A & 57042.985 & $95.89^{+1.19}_{-1.36}$ & $14.13^{+0.50}_{-0.50}$ & 113 & $0.34$ & ... (...)   \\
IC150127A & 57049.481 & $100.37^{+1.36}_{-1.63}$ & $4.59^{+0.79}_{-0.67}$ & 293 & $0.66$ & ... (...)   \\
IC150129A & 57051.227 & $358.51^{+3.91}_{-6.55}$ & $6.39^{+3.16}_{-3.67}$ & 130 & $0.33$ & 4FGL J2349.4+0534 (1.41)   \\
IC150224A & 57078.000 & $237.75^{+8.26}_{-2.26}$ & $55.11^{+3.38}_{-3.03}$ & 106 & $0.38$ & 4FGL J1553.1+5438 (0.56)   \\
IC150313A & 57094.321 & $127.05^{+1.76}_{-2.07}$ & $-3.36^{+0.75}_{-0.75}$ & 107 & $0.29$ & ... (...)   \\
IC150428A & 57140.591 & $31.07^{+4.04}_{-6.42}$ & $15.02^{+1.94}_{-1.42}$ & 109 & $0.32$ & 4FGL J0204.8+1513 (0.26)   \\
IC150515A & 57157.942 & $91.49^{+0.92}_{-0.75}$ & $12.14^{+0.53}_{-0.50}$ & 401 & $0.77$ & ... (...)   \\
IC150526A & 57168.017 & $139.79^{+2.46}_{-2.99}$ & $-1.49^{+0.90}_{-1.01}$ & 108 & $0.28$ & 4FGL J0914.1-0202 (1.36)   \\
IC150601A & 57174.018 & $333.37^{+2.42}_{-1.71}$ & $9.63^{+1.21}_{-1.17}$ & 106 & $0.27$ & ... (...)   \\
IC150609A & 57182.027 & $49.53^{+1.10}_{-1.10}$ & $0.30^{+0.45}_{-0.82}$ & 118 & $0.31$ & ... (...)   \\
IC150609B & 57182.180 & $245.43^{+1.67}_{-1.23}$ & $0.22^{+1.04}_{-0.93}$ & 116 & $0.30$ & 4FGL J1625.1-0020 (1.03)   \\
IC150625A & 57198.640 & $71.89^{+4.35}_{-4.70}$ & $0.86^{+2.39}_{-1.83}$ & 112 & $0.29$ & 4FGL J0442.6-0017 (1.69)   \\
IC150625B & 57198.732 & $306.43^{+2.02}_{-2.02}$ & $19.08^{+0.91}_{-1.18}$ & 154 & $0.46$ & 4FGL J2030.9+1935 (1.34)   \\
IC150714A & 57217.910 & $326.29^{+1.49}_{-1.32}$ & $26.36^{+1.89}_{-2.19}$ & 439 & $0.84$ & ... (...)   \\
IC150809A* & 57243.322 & $221.75^{+0.31}_{-0.26}$ & $-17.15^{+0.23}_{-0.16}$ & 11667 & $0.20$ & ... (...)   \\
IC150812A & 57246.318 & $317.59^{+5.10}_{-4.66}$ & $30.09^{+2.31}_{-2.43}$ & 125 & $0.44$ & 3FHL J2115.2+2933 (1.19)   \\
IC150812B & 57246.759 & $328.27^{+0.75}_{-0.88}$ & $6.17^{+0.49}_{-0.53}$ & 508 & $0.83$ & ... (...)   \\
IC150823A & 57257.623 & $325.90^{+3.47}_{-4.17}$ & $-2.35^{+2.61}_{-2.09}$ & 133 & $0.35$ & 4FGL J2148.9-0121 (1.66)   \\
IC150831A & 57265.218 & $54.76^{+0.92}_{-0.92}$ & $34.00^{+1.13}_{-1.21}$ & 181 & $0.58$ & ... (...)   \\
IC150904A & 57269.760 & $133.77^{+0.53}_{-0.88}$ & $28.08^{+0.51}_{-0.55}$ & 302 & $0.74$ & 3FHL J0854.1+2752 (0.29)   \\
IC150914A & 57279.875 & $129.68^{+1.89}_{-2.59}$ & $30.35^{+1.88}_{-1.29}$ & 120 & $0.43$ &    SWIFT J0840.2+29 (0.63)   \\
IC150918A & 57283.546 & $49.83^{+2.50}_{-3.74}$ & $-2.95^{+1.35}_{-1.34}$ & 105 & $0.28$ &    SWIFT J0324.9-03 (1.41)   \\
IC150919A & 57284.206 & $279.54^{+1.76}_{-2.29}$ & $30.35^{+2.19}_{-1.50}$ & 228 & $0.67$ & 4FGL J1836.4+3137 (1.32)   \\
IC150923A & 57288.027 & $103.23^{+0.70}_{-1.14}$ & $3.96^{+0.60}_{-0.75}$ & 216 & $0.33$ & ... (...)   \\
IC150926A & 57291.901 & $194.55^{+0.79}_{-1.23}$ & $-4.56^{+0.93}_{-0.64}$ & 216 & $0.30$ & 4FGL J1258.7-0452 (0.34)   \\
IC151013A & 57308.124 & $178.72^{+1.11}_{-1.15}$ & $52.37^{+1.11}_{-1.11}$ & 156 & $0.52$ & ... (...)   \\
IC151017A & 57312.676 & $197.53^{+2.46}_{-2.72}$ & $19.95^{+3.01}_{-2.29}$ & 321 & $0.75$ & 4FGL J1311.8+2057 (1.09)   \\
IC151114A & 57340.873 & $76.16^{+1.36}_{-1.36}$ & $12.71^{+0.65}_{-0.73}$ & 1124 & $0.96$ & ... (...)   \\
IC151122A & 57348.532 & $262.05^{+0.88}_{-1.05}$ & $-2.24^{+0.63}_{-0.67}$ & 253 & $0.64$ & ... (...)   \\
IC160104A & 57391.444 & $79.41^{+0.83}_{-0.75}$ & $5.00^{+0.86}_{-0.97}$ & 217 & $0.57$ & 4FGL J0515.9+0537 (0.75)   \\
IC160128A & 57415.183 & $263.76^{+1.10}_{-1.80}$ & $-14.90^{+1.08}_{-1.20}$ & 583 & $0.15$ & ... (...)   \\
IC160225A & 57443.880 & $311.87^{+2.18}_{-1.78}$ & $60.06^{+1.65}_{-1.37}$ & 188 & $0.60$ & ... (...)   \\
IC160307A & 57454.697 & $91.32^{+7.08}_{-8.66}$ & $10.47^{+2.74}_{-4.45}$ & 106 & $0.28$ & 4FGL J0608.6+1149 (1.59)   \\
IC160331A & 57478.565 & $151.22^{+0.66}_{-0.66}$ & $15.48^{+0.66}_{-0.73}$ & 492 & $0.85$ & ... (...)   \\
IC160410A & 57488.735 & $235.63^{+1.23}_{-1.45}$ & $-4.07^{+1.31}_{-0.86}$ & 131 & $0.37$ & ... (...)   \\
IC160427A & 57505.245 & $240.29^{+0.44}_{-0.48}$ & $9.71^{+0.57}_{-0.42}$ & 85 & $0.45$ & ... (...)   \\
IC160510A & 57518.664 & $352.88^{+1.76}_{-1.45}$ & $1.90^{+0.75}_{-0.67}$ & 208 & $0.39$ & ... (...)   \\
IC160612A & 57551.434 & $16.52^{+0.88}_{-0.18}$ & $4.67^{+1.87}_{-0.52}$ & 106 & $0.25$ & ... (...)   \\
IC160614A & 57553.526 & $214.76^{+3.16}_{-4.13}$ & $40.82^{+3.33}_{-3.98}$ & 112 & $0.41$ & 4FGL J1421.1+3859 (1.87)   \\
IC160615A & 57554.404 & $304.32^{+1.63}_{-1.05}$ & $12.64^{+1.33}_{-1.34}$ & 150 & $0.41$ & 4FGL J2014.9+1225 (0.61)   \\
IC160707A & 57576.168 & $351.43^{+1.54}_{-2.29}$ & $0.60^{+0.82}_{-1.12}$ & 110 & $0.28$ & 4FGL J2326.2+0113 (0.64)   \\
IC160720A & 57589.914 & $60.25^{+10.72}_{-8.88}$ & $29.23^{+5.32}_{-5.87}$ & 108 & $0.37$ & 4FGL J0358.1+2850 (0.74)   \\
IC160727A & 57596.344 & $113.12^{+1.93}_{-1.54}$ & $14.67^{+1.08}_{-1.12}$ & 105 & $0.30$ & ... (...)   \\
IC160731A & 57600.080 & $214.58^{+0.53}_{-0.57}$ & $-0.30^{+0.45}_{-0.67}$ & 98 & $0.44$ & ... (...)   \\
IC160731A & 57600.785 & $312.63^{+3.74}_{-3.21}$ & $20.07^{+2.56}_{-2.13}$ & 118 & $0.39$ & 4FGL J2043.9+2051 (1.73)   \\
IC160806A & 57606.515 & $122.78^{+0.88}_{-1.23}$ & $-0.71^{+0.56}_{-0.56}$ & 219 & $0.58$ & ... (...)   \\
IC160812A & 57612.684 & $86.99^{+15.29}_{-15.29}$ & $48.83^{+9.95}_{-10.00}$ & 160 & $0.53$ & 4FGL J0553.5+4810 (1.14)   \\
IC160814A & 57614.907 & $200.04^{+3.12}_{-2.68}$ & $-32.13^{+1.75}_{-1.24}$ & 263 & $0.61$ &    SWIFT J1325.2-32 (1.25)   \\
IC160924A & 57655.741 & $241.13^{+4.92}_{-5.89}$ & $1.34^{+3.40}_{-2.80}$ & 191 & $0.51$ & 4FGL J1608.4+0055 (1.07)   \\
IC161001A & 57662.439 & $192.57^{+2.50}_{-2.07}$ & $37.12^{+1.51}_{-2.49}$ & 204 & $0.64$ & 4FGL J1249.8+3707 (0.09)   \\
IC161012A & 57673.613 & $190.06^{+2.20}_{-4.04}$ & $-7.48^{+2.18}_{-2.98}$ & 759 & $0.25$ &    SWIFT J1239.6-05 (2.14)   \\
IC161021A & 57682.309 & $121.42^{+2.64}_{-2.90}$ & $23.72^{+1.93}_{-2.02}$ & 135 & $0.43$ & 4FGL J0803.0+2439 (1.12)   \\
IC161027A & 57688.570 & $119.00^{+2.94}_{-2.24}$ & $1.53^{+2.32}_{-2.39}$ & 155 & $0.38$ & ... (...)   \\
IC161103A & 57695.380 & $40.87^{+1.05}_{-0.57}$ & $12.52^{+1.15}_{-0.61}$ & 85 & $0.31$ & 4FGL J0244.7+1316 (0.82)   \\
IC161117A & 57709.332 & $78.66^{+1.85}_{-1.93}$ & $1.60^{+1.90}_{-1.79}$ & 190 & $0.50$ & ... (...)   \\
IC161125A & 57717.430 & $140.01^{+2.15}_{-1.19}$ & $-0.11^{+0.75}_{-0.86}$ & 161 & $0.40$ & ... (...)   \\
IC161127A & 57719.665 & $257.55^{+36.46}_{-29.23}$ & $73.27^{+5.71}_{-9.96}$ & 139 & $0.45$ & 4FGL J1651.6+7219 (1.66)   \\
IC161210A & 57732.838 & $46.36^{+2.37}_{-0.92}$ & $15.25^{+0.93}_{-1.08}$ & 80 & $0.38$ & ... (...)   \\
IC161224A & 57746.537 & $61.79^{+2.50}_{-2.37}$ & $17.78^{+1.46}_{-1.56}$ & 139 & $0.42$ &    SWIFT J0413.3+16 (1.69)   \\
IC170105A & 57758.142 & $309.95^{+5.01}_{-7.56}$ & $8.16^{+2.00}_{-3.34}$ & 198 & $0.54$ &    SWIFT J2033.1+09 (2.41)   \\
IC170206A & 57790.549 & $180.35^{+5.23}_{-3.82}$ & $33.20^{+1.85}_{-2.16}$ & 135 & $0.46$ & 4FGL J1205.8+3321 (0.94)   \\
IC170208A & 57792.128 & $99.67^{+2.59}_{-3.30}$ & $16.84^{+1.60}_{-1.55}$ & 151 & $0.43$ & 3HWC J0634+165 (1.14)   \\
IC170208A & 57792.595 & $92.81^{+1.23}_{-1.05}$ & $4.59^{+0.94}_{-1.16}$ & 133 & $0.33$ & ... (...)   \\
IC170227A & 57811.065 & $205.09^{+1.89}_{-3.96}$ & $4.26^{+1.09}_{-1.12}$ & 108 & $0.27$ &    SWIFT J1338.2+04 (0.59)   \\
IC170308A & 57820.925 & $155.35^{+2.02}_{-1.19}$ & $5.53^{+0.98}_{-0.90}$ & 107 & $0.25$ & 4FGL J1019.7+0511 (0.53)   \\
IC170321A & 57833.314 & $98.26^{+1.32}_{-0.92}$ & $-15.06^{+1.04}_{-1.20}$ & 231 & $0.24$ & ... (...)   \\
IC170422A & 57865.646 & $240.95^{+3.34}_{-5.71}$ & $5.53^{+0.83}_{-1.01}$ & 161 & $0.39$ & ... (...)   \\
IC170427A & 57870.314 & $5.32^{+4.48}_{-5.27}$ & $-0.60^{+1.75}_{-1.23}$ & 155 & $0.38$ & 3FHL J0022.0+0006 (0.73)   \\
IC170514A & 57887.175 & $311.97^{+2.20}_{-1.23}$ & $18.60^{+2.10}_{-1.10}$ & 109 & $0.34$ & ... (...)   \\
IC170514B & 57887.300 & $227.37^{+1.23}_{-1.10}$ & $30.65^{+1.40}_{-0.99}$ & 174 & $0.55$ & ... (...)   \\
IC170527A & 57900.070 & $178.59^{+2.77}_{-3.47}$ & $26.49^{+3.82}_{-3.45}$ & 124 & $0.42$ & 4FGL J1148.5+2629 (1.3)   \\
IC170621A & 57925.191 & $74.97^{+7.25}_{-7.78}$ & $25.08^{+5.57}_{-6.20}$ & 109 & $0.37$ &    SWIFT J0502.4+24 (0.68)   \\
IC170626A & 57930.519 & $280.99^{+3.03}_{-1.63}$ & $8.80^{+1.13}_{-0.90}$ & 201 & $0.55$ & 4FGL J1846.3+0919 (0.8)   \\
IC170704A & 57938.293 & $230.45^{+1.67}_{-1.71}$ & $23.36^{+1.10}_{-0.89}$ & 195 & $0.60$ & ... (...)   \\
IC170717A & 57951.818 & $208.39^{+1.67}_{-1.19}$ & $25.16^{+1.41}_{-1.35}$ & 534 & $0.87$ & ... (...)   \\
IC170803A & 57968.084 & $1.10^{+4.48}_{-1.76}$ & $4.63^{+0.41}_{-0.41}$ & 214 & $0.56$ & ... (...)   \\
IC170809A & 57974.597 & $21.27^{+0.75}_{-1.05}$ & $-2.28^{+0.60}_{-0.67}$ & 226 & $0.60$ & ... (...)   \\
IC170819A & 57984.276 & $26.98^{+1.85}_{-3.03}$ & $18.88^{+1.11}_{-1.10}$ & 167 & $0.51$ & ... (...)   \\
IC170824A & 57989.554 & $41.92^{+3.03}_{-3.56}$ & $12.37^{+1.46}_{-1.30}$ & 175 & $0.49$ &    SWIFT J0248.3+12 (0.38)   \\
IC170922A & 58018.871 & $77.43^{+1.14}_{-0.75}$ & $5.79^{+0.64}_{-0.41}$ & 264 & $0.63$ & 3FHL J0509.4+0542 (0.11)   \\
IC170923A & 58019.021 & $173.45^{+2.37}_{-2.55}$ & $-2.54^{+0.90}_{-1.31}$ & 202 & $0.56$ & ... (...)   \\
IC171006A & 58032.308 & $132.63^{+1.41}_{-2.24}$ & $17.23^{+1.06}_{-0.66}$ & 118 & $0.37$ & ... (...)   \\
IC171015A & 58041.066 & $162.91^{+2.99}_{-1.71}$ & $-15.48^{+1.62}_{-1.98}$ & 72 & $0.55$ &   SWIFT J1051.2-170 (1.58)   \\
IC171028A & 58054.765 & $294.52^{+3.56}_{-3.38}$ & $2.05^{+2.20}_{-3.21}$ & 133 & $0.34$ & 3FHL J1927.5+0153 (2.64)   \\
IC171106A & 58063.778 & $340.14^{+0.62}_{-0.62}$ & $7.44^{+0.30}_{-0.26}$ & 1573 & $0.97$ & ... (...)   \\
IC171108A* & 58065.755 & $269.65^{+0.22}_{-0.18}$ & $-20.70^{+0.16}_{-0.16}$ & 20310 & $0.18$ & ... (...)   \\
IC180117A & 58135.752 & $206.10^{+1.19}_{-1.14}$ & $3.92^{+0.71}_{-0.78}$ & 85 & $0.42$ & ... (...)   \\
IC180123A & 58141.677 & $77.12^{+2.50}_{-2.90}$ & $8.01^{+0.41}_{-0.49}$ & 416 & $0.79$ & ... (...)   \\
IC180125A & 58143.976 & $207.51^{+1.01}_{-0.57}$ & $23.77^{+0.57}_{-0.57}$ & 110 & $0.36$ & ... (...)   \\
IC180205A & 58154.004 & $17.40^{+1.36}_{-0.92}$ & $-10.54^{+0.76}_{-0.72}$ & 113 & $0.23$ & ... (...)   \\
IC180213A & 58162.378 & $66.97^{+2.46}_{-2.59}$ & $6.09^{+1.95}_{-1.72}$ & 111 & $0.27$ & 4FGL J0427.3+0504 (1.02)   \\
IC180228A & 58177.572 & $294.79^{+1.85}_{-1.71}$ & $26.40^{+0.79}_{-1.12}$ & 124 & $0.42$ & ... (...)   \\
IC180313A & 58190.679 & $287.18^{+0.75}_{-2.46}$ & $5.53^{+0.34}_{-0.26}$ & 160 & $0.39$ & ... (...)   \\
IC180314A & 58191.804 & $58.71^{+1.89}_{-1.67}$ & $0.78^{+1.01}_{-1.01}$ & 145 & $0.36$ & ... (...)   \\
IC180316A & 58193.243 & $271.71^{+1.19}_{-3.43}$ & $-1.42^{+1.23}_{-1.27}$ & 156 & $0.39$ & 4FGL J1759.0-0107 (1.97)   \\
IC180410A & 58218.777 & $218.50^{+0.79}_{-1.27}$ & $0.56^{+0.75}_{-0.71}$ & 234 & $0.60$ & ... (...)   \\
IC180417A & 58225.279 & $305.73^{+3.60}_{-1.58}$ & $-4.41^{+0.67}_{-0.75}$ & 202 & $0.58$ & ... (...)   \\
IC180528A & 58266.506 & $312.14^{+1.41}_{-2.02}$ & $0.30^{+0.86}_{-1.45}$ & 110 & $0.28$ & 4FGL J2049.7-0036 (0.97)   \\
IC180608A & 58277.597 & $69.08^{+1.63}_{-1.41}$ & $-1.08^{+0.78}_{-0.78}$ & 158 & $0.40$ & 4FGL J0436.2-0038 (0.43)   \\
IC180612A & 58281.190 & $338.69^{+5.10}_{-5.71}$ & $3.73^{+2.81}_{-3.70}$ & 107 & $0.25$ &    SWIFT J2235.7+01 (2.12)   \\
IC180613A & 58282.982 & $38.06^{+5.84}_{-4.26}$ & $11.53^{+4.15}_{-4.91}$ & 155 & $0.41$ & 4FGL J0231.8+1322 (1.84)   \\
IC180728A* & 58327.845 & $74.14^{+0.44}_{-0.35}$ & $-17.74^{+0.51}_{-0.59}$ & 16952 & $0.18$ & ... (...)   \\
IC180807A & 58337.202 & $100.37^{+4.00}_{-5.05}$ & $11.15^{+2.98}_{-2.12}$ & 106 & $0.28$ & 4FGL J0642.4+1048 (0.41)   \\
IC180908A & 58369.833 & $144.98^{+1.49}_{-2.20}$ & $-2.39^{+1.16}_{-1.12}$ & 144 & $0.30$ & ... (...)   \\
IC180909A & 58370.604 & $141.37^{+1.05}_{-1.27}$ & $26.94^{+0.88}_{-1.00}$ & 171 & $0.53$ & ... (...)   \\
IC180919A & 58380.065 & $258.40^{+1.49}_{-1.49}$ & $32.84^{+0.94}_{-0.75}$ & 144 & $0.48$ & 4FGL J1714.6+3228 (0.43)   \\
IC181008A & 58399.779 & $77.08^{+2.68}_{-3.56}$ & $1.23^{+1.23}_{-1.16}$ & 108 & $0.27$ & ... (...)   \\
IC181014A & 58405.495 & $225.22^{+1.36}_{-2.64}$ & $-34.95^{+1.22}_{-1.79}$ & 62 & $0.39$ & 4FGL J1505.0-3433 (0.94)   \\
IC181023A & 58414.693 & $270.18^{+1.89}_{-1.71}$ & $-8.42^{+1.13}_{-1.55}$ & 237 & $0.15$ & 4FGL J1804.4-0852 (1.03)   \\
IC181023B & 58414.736 & $78.27^{+1.76}_{-0.92}$ & $21.54^{+0.96}_{-0.93}$ & 136 & $0.43$ & ... (...)   \\
IC181114A & 58436.945 & $6.02^{+1.63}_{-2.24}$ & $18.84^{+0.87}_{-0.98}$ & 145 & $0.44$ & ... (...)   \\
IC181120A & 58442.709 & $25.71^{+5.54}_{-5.27}$ & $11.72^{+2.41}_{-4.50}$ & 188 & $0.54$ & 4FGL J0150.9+1230 (2.13)   \\
IC181120B & 58442.944 & $324.58^{+7.74}_{-9.04}$ & $51.74^{+6.75}_{-9.48}$ & 173 & $0.57$ &    SWIFT J2133.6+51 (0.94)   \\
IC181121A & 58443.580 & $132.19^{+7.34}_{-6.99}$ & $32.93^{+4.19}_{-3.57}$ & 209 & $0.65$ &    SWIFT J0848.1+34 (1.81)   \\
IC181212A & 58464.085 & $316.41^{+1.85}_{-2.02}$ & $-31.00^{+1.68}_{-1.58}$ & 162 & $0.46$ & 4FGL J2112.5-3043 (1.51)   \\
IC190113A & 58496.089 & $56.91^{+1.63}_{-1.41}$ & $-0.82^{+0.75}_{-0.82}$ & 156 & $0.39$ & ... (...)   \\
IC190124A & 58507.155 & $307.44^{+0.53}_{-1.14}$ & $-32.22^{+0.96}_{-0.31}$ & 157 & $0.74$ & ... (...)   \\
IC190201A & 58515.016 & $245.08^{+0.75}_{-0.88}$ & $38.78^{+0.77}_{-0.67}$ & 163 & $0.53$ & ... (...)   \\
IC190214A & 58528.673 & $228.25^{+0.79}_{-0.53}$ & $-4.14^{+0.37}_{-0.30}$ & 348 & $0.74$ & ... (...)   \\
IC190221A & 58535.351 & $268.59^{+1.41}_{-1.58}$ & $-17.00^{+1.24}_{-0.51}$ & 56 & $0.55$ & ... (...)   \\
IC190223A & 58537.850 & $155.21^{+0.70}_{-0.66}$ & $19.67^{+0.28}_{-0.44}$ & 168 & $0.51$ & ... (...)   \\
IC190317A & 58559.832 & $81.25^{+5.89}_{-5.98}$ & $3.21^{+3.93}_{-4.07}$ & 108 & $0.26$ & 3FHL J0521.6+0104 (2.3)   \\
IC190410A & 58583.436 & $310.61^{+3.30}_{-3.65}$ & $12.22^{+2.84}_{-2.28}$ & 105 & $0.28$ & 4FGL J2044.0+1036 (1.66)   \\
IC190413A & 58586.450 & $219.33^{+0.70}_{-1.32}$ & $11.72^{+0.72}_{-0.72}$ & 107 & $0.29$ & 4FGL J1438.6+1205 (0.49)   \\
IC190413B & 58586.665 & $245.57^{+1.23}_{-1.49}$ & $21.98^{+1.21}_{-1.44}$ & 115 & $0.38$ & ... (...)   \\
IC190415A & 58588.437 & $154.86^{+2.94}_{-4.70}$ & $5.27^{+2.48}_{-1.95}$ & 117 & $0.30$ & 4FGL J1019.7+0511 (0.11)   \\
IC190422A & 58595.250 & $166.90^{+3.21}_{-3.03}$ & $17.39^{+2.00}_{-2.56}$ & 170 & $0.51$ & 4FGL J1112.4+1751 (1.24)   \\
IC190503A & 58606.724 & $120.19^{+0.66}_{-0.66}$ & $6.43^{+0.68}_{-0.75}$ & 142 & $0.34$ & ... (...)   \\
IC190504A & 58607.768 & $65.17^{+1.67}_{-1.14}$ & $-37.26^{+0.61}_{-1.09}$ & 55 & $0.39$ & 3FHL J0420.4-3744 (0.48)   \\
IC190515A & 58618.451 & $127.88^{+0.79}_{-0.83}$ & $12.60^{+0.50}_{-0.46}$ & 457 & $0.82$ & ... (...)   \\
IC190613A & 58647.829 & $312.19^{+0.66}_{-0.79}$ & $26.57^{+0.75}_{-0.71}$ & 195 & $0.61$ & ... (...)   \\
IC190619A & 58653.552 & $343.52^{+4.13}_{-3.16}$ & $10.28^{+2.02}_{-2.76}$ & 199 & $0.55$ &   SWIFT J2254.2+114 (1.49)   \\
IC190629A & 58663.809 & $29.12^{+39.68}_{-118.65}$ & $84.56^{+4.66}_{-4.40}$ & 109 & $0.34$ & 3FHL J0249.7+8434 (1.26)   \\
IC190704A & 58668.784 & $161.81^{+2.15}_{-3.91}$ & $26.90^{+1.94}_{-1.91}$ & 155 & $0.49$ & 4FGL J1049.8+2741 (0.97)   \\
IC190712A & 58676.052 & $76.64^{+5.23}_{-6.99}$ & $12.75^{+4.79}_{-2.82}$ & 109 & $0.30$ & 4FGL J0502.5+1340 (1.34)   \\
IC190730A & 58694.869 & $226.14^{+1.27}_{-1.98}$ & $10.77^{+1.03}_{-1.17}$ & 298 & $0.67$ & 3FHL J1504.3+1030 (0.27)   \\
IC190819A & 58714.732 & $148.54^{+2.29}_{-3.30}$ & $1.45^{+0.93}_{-0.75}$ & 113 & $0.29$ & 3FHL J0946.2+0104 (2.01)   \\
IC190922A & 58748.405 & $167.30^{+2.81}_{-2.72}$ & $-22.27^{+3.39}_{-3.31}$ & 3114 & $0.20$ & 3FHL J1103.6-2328 (1.77)   \\
IC190922B & 58748.961 & $5.71^{+1.19}_{-1.27}$ & $-1.53^{+0.90}_{-0.78}$ & 187 & $0.50$ & ... (...)   \\
IC191001A & 58757.840 & $313.99^{+6.94}_{-2.46}$ & $12.79^{+1.65}_{-1.64}$ & 218 & $0.59$ & 4FGL J2052.7+1218 (0.91)   \\
IC191119A & 58806.043 & $229.31^{+5.49}_{-4.97}$ & $3.77^{+2.47}_{-2.24}$ & 177 & $0.45$ & 4FGL J1521.1+0421 (1.15)   \\
IC191122A & 58809.948 & $27.03^{+1.98}_{-2.72}$ & $0.07^{+1.08}_{-1.57}$ & 127 & $0.33$ & ... (...)   \\
IC191204A & 58821.949 & $80.16^{+2.42}_{-1.98}$ & $2.87^{+1.05}_{-0.97}$ & 130 & $0.33$ & ... (...)   \\
IC191215A & 58832.465 & $286.83^{+2.27}_{-1.92}$ & $58.45^{+2.08}_{-1.86}$ & 133 & $0.48$ & ... (...)   \\
IC191231A & 58848.458 & $48.47^{+5.98}_{-7.65}$ & $20.11^{+4.48}_{-3.73}$ & 156 & $0.46$ & 4FGL J0312.7+2012 (0.28)   \\
IC200109A & 58857.987 & $165.45^{+3.60}_{-4.39}$ & $11.80^{+1.18}_{-1.29}$ & 375 & $0.77$ & 3FHL J1103.1+1156 (0.35)   \\
IC200117A & 58865.464 & $116.02^{+0.79}_{-1.19}$ & $29.18^{+0.86}_{-0.81}$ & 108 & $0.38$ &    SWIFT J0744.0+29 (0.09)   \\
IC200120A* & 58868.784 & $67.41^{+0.40}_{-0.31}$ & $-14.59^{+0.23}_{-0.27}$ & 6055 & $0.31$ & ... (...)   \\
IC200410A & 58949.972 & $242.58^{+10.20}_{-10.20}$ & $11.61^{+7.83}_{-6.19}$ & 110 & $0.30$ &    SWIFT J1608.8+12 (0.78)   \\
IC200421A & 58960.025 & $87.93^{+3.43}_{-2.81}$ & $8.23^{+2.08}_{-1.81}$ & 127 & $0.33$ & ... (...)   \\
IC200425A & 58964.977 & $99.97^{+4.76}_{-3.00}$ & $53.72^{+2.25}_{-1.69}$ & 135 & $0.48$ &    SWIFT J0645.9+53 (1.14)   \\
IC200512A & 58981.314 & $295.18^{+1.67}_{-2.24}$ & $15.79^{+1.24}_{-1.28}$ & 109 & $0.32$ & ... (...)   \\
IC200523A & 58992.104 & $338.64^{+9.98}_{-6.02}$ & $1.75^{+1.79}_{-3.51}$ & 105 & $0.25$ &    SWIFT J2235.7+01 (0.3)   \\
IC200530A & 58999.330 & $255.37^{+2.46}_{-2.55}$ & $26.61^{+2.32}_{-3.25}$ & 82 & $0.59$ & 4FGL J1702.2+2642 (0.2)   \\
IC200614A & 59014.529 & $33.84^{+4.79}_{-6.37}$ & $31.61^{+2.71}_{-2.25}$ & 115 & $0.41$ & 4FGL J0220.2+3246 (1.57)   \\
IC200615A & 59015.618 & $142.95^{+1.14}_{-1.41}$ & $3.66^{+1.16}_{-1.01}$ & 496 & $0.83$ & ... (...)   \\
IC200620A & 59020.127 & $162.11^{+0.62}_{-0.92}$ & $11.95^{+0.61}_{-0.46}$ & 114 & $0.33$ & ... (...)   \\
IC200806A & 59067.577 & $157.25^{+1.17}_{-0.87}$ & $47.75^{+0.64}_{-0.59}$ & 107 & $0.40$ & ... (...)   \\
IC200911A & 59103.597 & $51.11^{+4.39}_{-10.99}$ & $38.11^{+2.31}_{-1.97}$ & 111 & $0.41$ &    SWIFT J0333.3+37 (1.94)   \\
IC200916A & 59108.861 & $109.78^{+1.05}_{-1.41}$ & $14.36^{+0.85}_{-0.81}$ & 110 & $0.32$ & ... (...)   \\
IC200921A & 59113.797 & $195.29^{+2.33}_{-1.71}$ & $26.24^{+1.46}_{-1.73}$ & 117 & $0.41$ & 3FHL J1303.0+2435 (1.7)   \\
IC200926A & 59118.329 & $96.46^{+0.70}_{-0.53}$ & $-4.33^{+0.60}_{-0.75}$ & 670 & $0.44$ & ... (...)   \\
IC200926B & 59118.941 & $184.75^{+3.65}_{-1.54}$ & $32.93^{+1.16}_{-0.88}$ & 121 & $0.43$ & ... (...)   \\
IC200929A & 59121.742 & $29.53^{+0.53}_{-0.53}$ & $3.47^{+0.71}_{-0.34}$ & 183 & $0.47$ & ... (...)   \\
IC201007A & 59129.918 & $265.17^{+0.48}_{-0.48}$ & $5.34^{+0.30}_{-0.19}$ & 683 & $0.89$ & ... (...)   \\
IC201014A & 59136.093 & $221.22^{+0.97}_{-1.19}$ & $14.44^{+0.66}_{-0.46}$ & 147 & $0.41$ & ... (...)   \\
IC201021A & 59143.276 & $260.82^{+1.71}_{-1.67}$ & $14.55^{+1.31}_{-0.69}$ & 105 & $0.30$ & ... (...)   \\
IC201114A & 59167.629 & $105.73^{+0.92}_{-1.27}$ & $5.87^{+1.05}_{-1.01}$ & 214 & $0.56$ & ... (...)   \\
IC201115A & 59168.088 & $195.12^{+1.23}_{-1.45}$ & $1.38^{+1.27}_{-1.08}$ & 177 & $0.46$ & ... (...)   \\
IC201120A & 59173.406 & $307.66^{+5.19}_{-5.68}$ & $40.72^{+5.02}_{-2.75}$ & 154 & $0.50$ & 4FGL J2032.6+4053 (0.41)   \\
IC201130A & 59183.848 & $30.54^{+1.10}_{-1.27}$ & $-12.10^{+1.14}_{-1.11}$ & 203 & $0.15$ & 4FGL J0206.4-1151 (1.07)   \\
IC201209A & 59192.428 & $6.86^{+1.01}_{-1.19}$ & $-9.25^{+0.94}_{-1.10}$ & 419 & $0.19$ & ... (...)   \\
IC201221A & 59204.526 & $261.69^{+2.28}_{-2.46}$ & $41.81^{+1.25}_{-1.14}$ & 175 & $0.56$ & ... (...)   \\
IC201222A & 59205.039 & $206.37^{+0.88}_{-0.75}$ & $13.44^{+0.54}_{-0.34}$ & 186 & $0.53$ & ... (...)   \\

    \enddata
  \end{deluxetable*}


\section*{Acknowledgements}
The IceCube collaboration acknowledges the significant contributions to this manuscript from Mehr Un Nisa. We also acknowledge support from:
USA {\textendash} U.S. National Science Foundation-Office of Polar Programs,
U.S. National Science Foundation-Physics Division,
U.S. National Science Foundation-EPSCoR,
Wisconsin Alumni Research Foundation,
Center for High Throughput Computing (CHTC) at the University of Wisconsin{\textendash}Madison,
Open Science Grid (OSG),
Advanced Cyberinfrastructure Coordination Ecosystem: Services {\&} Support (ACCESS),
Frontera computing project at the Texas Advanced Computing Center,
U.S. Department of Energy-National Energy Research Scientific Computing Center,
Particle astrophysics research computing center at the University of Maryland,
Institute for Cyber-Enabled Research at Michigan State University,
and Astroparticle physics computational facility at Marquette University;
Belgium {\textendash} Funds for Scientific Research (FRS-FNRS and FWO),
FWO Odysseus and Big Science programmes,
and Belgian Federal Science Policy Office (Belspo);
Germany {\textendash} Bundesministerium f{\"u}r Bildung und Forschung (BMBF),
Deutsche Forschungsgemeinschaft (DFG),
Helmholtz Alliance for Astroparticle Physics (HAP),
Initiative and Networking Fund of the Helmholtz Association,
Deutsches Elektronen Synchrotron (DESY),
and High Performance Computing cluster of the RWTH Aachen;
Sweden {\textendash} Swedish Research Council,
Swedish Polar Research Secretariat,
Swedish National Infrastructure for Computing (SNIC),
and Knut and Alice Wallenberg Foundation;
European Union {\textendash} EGI Advanced Computing for research;
Australia {\textendash} Australian Research Council;
Canada {\textendash} Natural Sciences and Engineering Research Council of Canada,
Calcul Qu{\'e}bec, Compute Ontario, Canada Foundation for Innovation, WestGrid, and Compute Canada;
Denmark {\textendash} Villum Fonden, Carlsberg Foundation, and European Commission;
New Zealand {\textendash} Marsden Fund;
Japan {\textendash} Japan Society for Promotion of Science (JSPS)
and Institute for Global Prominent Research (IGPR) of Chiba University;
Korea {\textendash} National Research Foundation of Korea (NRF);
Switzerland {\textendash} Swiss National Science Foundation (SNSF);
United Kingdom {\textendash} Department of Physics, University of Oxford.
\clearpage
\appendix

\begin{figure}[h]
\includegraphics[width=1\textwidth, trim=0cm 0cm 0cm 0cm, clip=true]{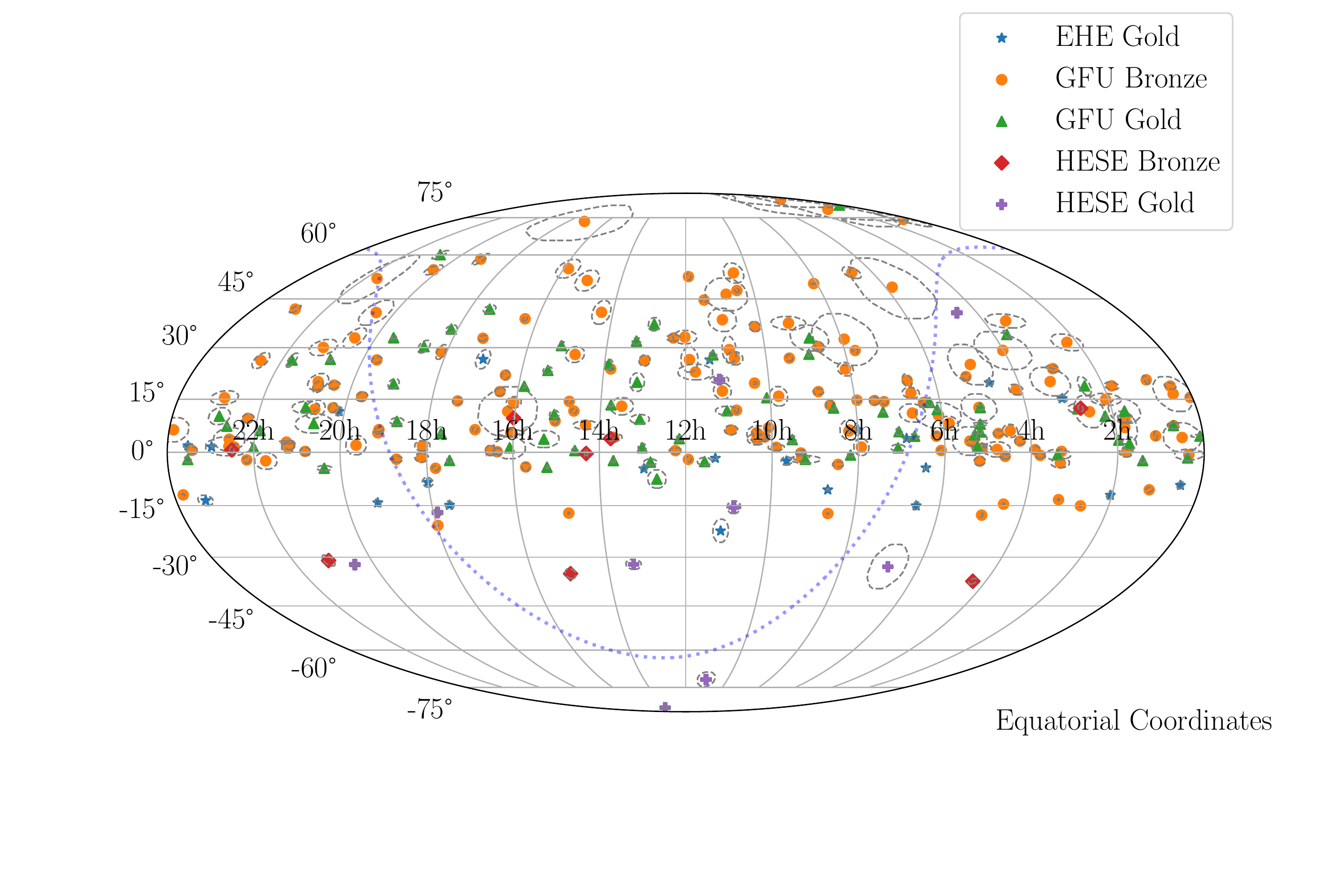}
\caption{The all-sky distribution of the alerts in the catalog in equatorial coordinates. The blue stars denote EHE, the orange circles GFU Bronze, the green triangles shows GFU Gold, the red diamonds show HESE Bronze, and the purple plus-signs show HESE Gold alerts.The 90\% uncertainty contours at the location of each alert are shown by the dashed ellipses.}
\label{fig:map2}
\end{figure}

\begin{figure}[h]
\includegraphics[width=1\textwidth]{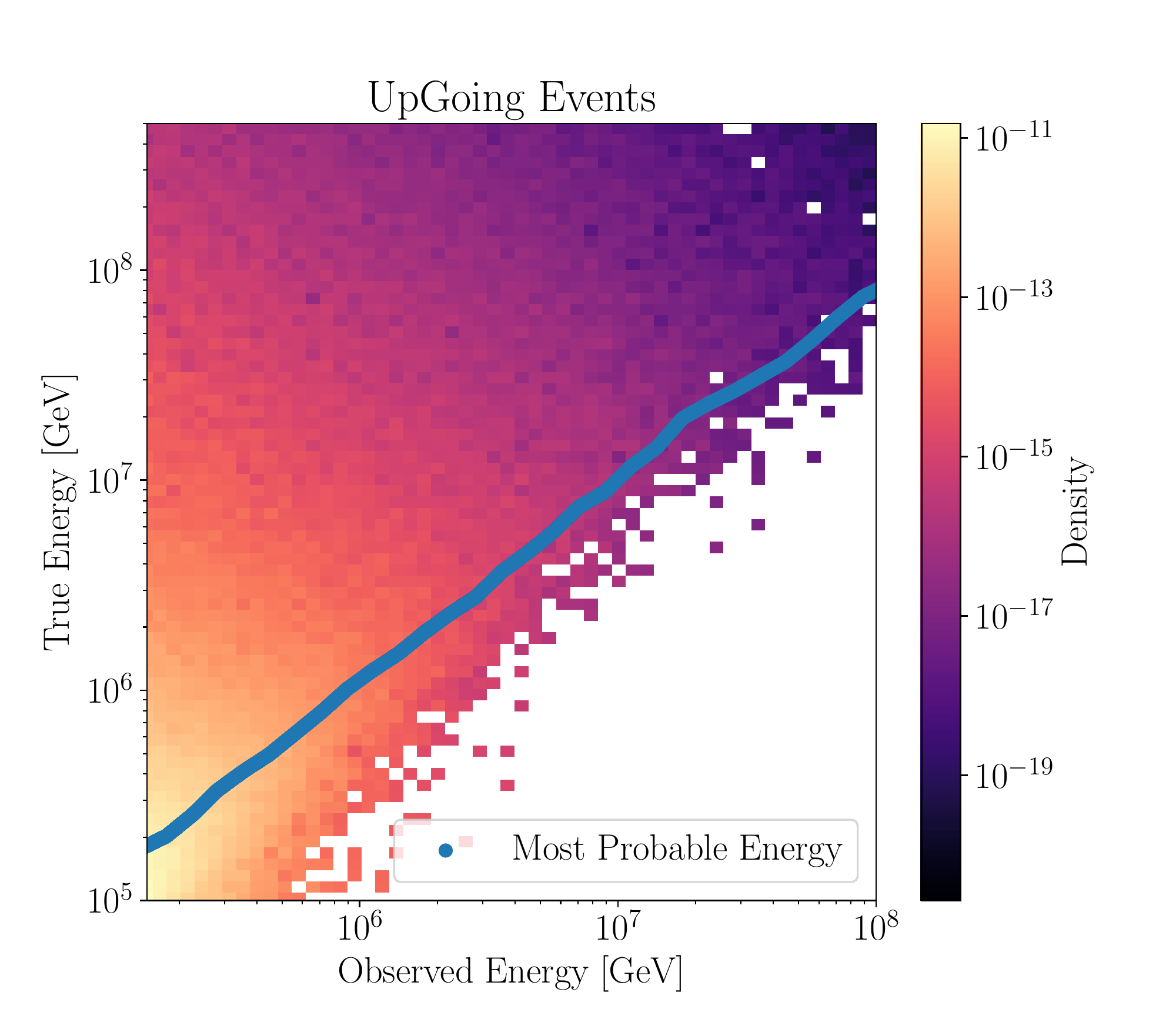}
\caption{The true neutrino energy as a function of observed energy for simulated up-going GFU Gold events. The most likely estimated neutrino energy is shown in blue.}
\label{fig:energy_mc}
\end{figure}
\clearpage

\bibliography{main}{}
\bibliographystyle{aasjournal}

\end{document}